\documentclass[prb,letterpaper,aps,floatfix,twocolumn]{revtex4-1}
\usepackage{graphicx}
\usepackage{amsmath}
\usepackage[small,bf]{subfigure}
\newcommand{\beq}{\begin{equation}}
\newcommand{\eeq}{\end{equation}}
\newcommand{\bk}{{{\bf{k}}}}

\newcommand{\br}{{{\bf{r}}}}

\newcommand{\bA}{{\bf{A}}}
\newcommand{\bB}{{\bf{B}}}

\newcommand{\bq}{{\bf{q}}}

\newcommand{\beqa}{\begin{eqnarray}}
\newcommand{\eeqa}{\end{eqnarray}}
\newcommand{\pdg}{{\vphantom \dag}}
\newcommand{\dg}{{\dag}}
\newcommand{\bnabla}{{\boldsymbol \nabla}}

\newcommand{\bGamma}{{\boldsymbol \Gamma}}
\newcommand{\upa}{\uparrow}
\newcommand{\da}{\downarrow} 
\newcommand{\cG}{{\cal G}}

\newcommand{\cI}{{\cal I}}

\newcommand{\cD}{{\cal D}}
\newcommand{\cP}{{\cal P}}
\begin{document}
\title{Dynamical density response and optical conductivity in topological metals}
\author{A.A. Burkov}
\affiliation{Department of Physics and Astronomy, University of Waterloo, Waterloo, Ontario 
N2L 3G1, Canada} 
\date{\today}
\begin{abstract}
Topological metals continue to attract attention as novel gapless states of matter. 
While there by now exists an exhaustive classification of possible topologically nontrivial metallic states, 
their observable properties, that follow from the electronic structure topology, are less well understood. 
Here we present a study of the electromagnetic response of three-dimensional topological metals
with Weyl or Dirac nodes in the spectrum, which systematizes and extends earlier pioneering studies. 
In particular, we argue that a smoking-gun feature of the chiral anomaly in topological metals 
is the existence of propagating chiral density modes even in the regime of weak magnetic fields. 
We also demonstrate that the optical conductivity of such metals exhibits an
extra peak, which exists on top of the standard metallic Drude peak. The spectral weight of this peak is transferred from high frequencies 
and its width is proportional to the chiral charge relaxation rate.
\end{abstract}
\maketitle
\section{Introduction}
\label{sec:1}
Topological metal (TM) is a recently discovered new phase of matter.~\cite{Weyl_RMP,Hasan_ARCMP,Felser_ARCMP,Burkov_ARCMP,Wan11,Burkov11-1,Burkov11-2,Xu11,Kane12,Fang12,Fang13,Chen14,Neupane14,HasanTaAs,DingTaAs2,DingTaAs,Lu15,Felser17}
It is characterized by topological invariants, defined on the Fermi surface,~\cite{Volovik03,Haldane04,Volovik07,Murakami07}
rather than in the whole Brillouin zone (BZ), as in topological insulators (TI). 
Such Fermi surface topological invariants arise as a consequence of monopole-like singularities in the electronic structure, 
Weyl nodes, whose significance was emphasized early on by Volovik and by Murakami.~\cite{Volovik03,Murakami07}

Perhaps the most interesting feature of TM is that their electronic structure topology leads not only to spectroscopic manifestations in the form of edge states,~\cite{Wan11} a feature they share with TI, but also to nontrivial response. 
This novel response is usually described as being a consequence of the chiral anomaly,~\cite{Zyuzin12-1} which may be understood in the following way.
While the appearance of gapless Weyl  nodes in the spectrum
has a topological origin, it also leads to an emergent symmetry, or an emergent conservation law, namely conservation of the chiral charge. This conservation law becomes increasingly more precise as the Fermi energy of the TM approaches the Weyl nodes. However, this apparent low-energy conservation law is violated when the system is coupled to an electromagnetic field. 
The origin of this violation lies in the fact that the chiral symmetry can never be an exact symmetry of a $(3+1)$-dimensional Dirac 
fermion on a lattice, as first pointed out by Nielsen and Ninomiya,~\cite{Nielsen83} as a single (or, more generally, an odd number)
Dirac point in the BZ is topologically incompatible with the chiral symmetry.
Thus, while the chiral symmetry appears to be present when one focuses only on states at the small Fermi surface, enclosing the 
Weyl points, the global lack of chiral symmetry manifests in the electromagnetic response of the system. 
This property is of great interest both because it has a topological origin and because it is contrary to one of the fundamental postulates of the standard theory of metals, which states that anything of observable consequence in a metal involves only states 
on the Fermi surface. 

The chiral anomaly in TM has numerous predicted observable consequences, 
which include negative longitudinal magnetoresistance (LMR),~\cite{Spivak12,Burkov_lmr_prb} giant planar Hall 
effect (PHE),~\cite{Burkov_gphe,Tewari_gphe} and anomalous Hall effect.~\cite{Burkov_AHE}
While most of these have already been observed experimentally in various TM materials,~\cite{Ong_anomaly,Li_anomaly,Shen_gphe,Felser_gphe,Zhang_gphe,Felser17}
none of these phenomena by themselves may be regarded as smoking-gun manifestations of the chiral anomaly, 
in the sense that all of them may in principle arise from unrelated sources, and these sources all have to be ruled out before the chiral anomaly origin may be claimed. 
An excellent discussion of these issues in the case of the negative LMR may be found in Ref.~\onlinecite{Ong18}. 
 
As first discussed by Altland and Bagrets,~\cite{Altland15} a truly unique feature of the chiral anomaly is the highly unusual dependence of the transport properties, such as the sample conductance, on the relevant length (and time or frequency, as will be shown in this paper) scales. 
In an ordinary three-dimensional (3D) metal the conductance scales linearly with the sample size $L$
\beq
\label{eq:1}
G(L) = \sigma L, 
\eeq
where the Drude conductivity $\sigma$ is related to the density of states at the Fermi energy $g$ and the 
diffusion constant $D$ by the Einstein relation
\beq
\label{eq:2}
\sigma = e^2 g D.
\eeq
Corrections to Eq.~\eqref{eq:1} are small in good metals, the small parameter being $1/ k_F \ell$, where $k_F$ is the Fermi momentum 
($\hbar =  c = k_B = 1$ units are used henceforth) and $\ell$ is the mean free path; the corrections arise only at very low temperatures as a result of quantum interference phenomena. 
The scaling of Eq.~\eqref{eq:1} is partly a consequence of the fact that, in an ordinary metal in the diffusive transport regime, i.e. at length scales, longer than the mean free path $\ell$ and time scales longer than the momentum relaxation time $\tau$, no intrinsic 
hydrodynamic (i.e. long) length scales remain, besides the sample size $L$. 

However, as discussed in Ref.~\onlinecite{Altland15}, in a TM two additional hydrodynamic length scales 
emerge. These are the chiral charge diffusion length 
\beq
\label{eq:3}
L_c = \sqrt{D \tau_c}, 
\eeq
where $\tau_c \gg \tau$ is the chirality relaxation time, and 
\beq
\label{eq:4}
L_a = D/\Gamma, 
\eeq
where $\Gamma = e B/2 \pi^2 g$ and $B$ is the applied magnetic field. 
$L_a$ is a new purely quantum mechanical magnetic-field-related length scale, which is 
distinct from the magnetic length $\ell_B = 1/\sqrt{e B}$ and which arises from the chiral anomaly. 
It is related to the magnetic length as $L_a \sim \ell (k_F \ell_B)^2$ and is thus much longer than the mean free 
path in the weak-field (quasiclassical) regime $k_F \ell_B \gg 1$, which we will be interested in here. 
Transport properties of TM may then be shown to depend strongly on the interplay of the three 
length scales: $L$, $L_c$, and $L_a$.~\cite{Altland15,Burkov_gphe}
In particular, the strength of the negative LMR and the PHE depends on the parameter $L_c/L_a$, getting stronger as this ratio
increases. 

Particularly striking phenomena arise when $L_a < L < L_c^2/L_a$,~\cite{Altland15} which is an extended and accessible range when 
$L_c/L_a \gg 1$. 
In this regime the sample conductance is given by
\beq
\label{eq:5}
G(L) = \frac{e^2 N_{\phi}}{2 \pi}, 
\eeq
where $N_{\phi} = L^2/ 2 \pi \ell_B^2$ is the number of magnetic flux quanta, piercing the sample with cross-section area $L^2$. 
This means that in the regime $L_a < L < L_c^2/L_a$ the sample transports electric current as $N_{\phi}$ one-dimensional (1D)
conduction channels and the conduction is ballistic and dissipationless [of course Eq.~\eqref{eq:5} only represents the dominant part of the
conductance and ordinary dissipative Ohmic conduction is also present]. 
This is striking because it arises in a 3D metal with a Fermi surface and in the weak magnetic field regime $k_F \ell_B \gg 1$. 
The existence of such ballistic quasi-1D transport regime is a smoking-gun manifestation of the chiral anomaly 
in 3D TM. 

In this paper we further elaborate on this striking property of TM and consider their related dynamical properties. 
In particular, we demonstrate that the quasi-1D transport regime manifests in dynamics as chiral propagating density modes, 
which exist in a range of wavevector values given by
\beq
\label{eq:6}
L_a/L_c^2 < q < 1/L_a. 
\eeq
This ``one-dimensionalization" of the electron dynamics is a unique property of TM, related to the chiral anomaly. 

We also demonstrate that related phenomena exist in frequency-dependent properties of TM. 
In particular we demonstrate that the frequency dependence of the optical conductivity of TM has a non-Drude form, 
where an extra narrow peak exists at low frequencies, whose width scales as $1/\tau_c$ while height is a function of the ratio $L_c/L_a$. 
The spectral weight of this extra peak is transferred from high frequencies. 

The rest of the paper is organized as follows. 
In Section~\ref{sec:2} we calculate the full density response function of a simple model of a TM in an external 
magnetic field. We analyze the eigenmode structure of the density response function and demonstrate the presence of chiral 
propagating density modes when $L_a/L_c^2 < q < 1/L_a$. 
In Section~\ref{sec:3} we relate the existence of these propagating chiral modes to observable transport properties of TM. 
We also demonstrate that similar phenomena exist in the frequency domain: we analyze the frequency dependence of the optical conductivity and point out its non-Drude nature. 
We conclude in Section~\ref{sec:4} with a brief discussion of the main results. 

\section{Density response function of a topological metal}
\label{sec:2}
We start from the simplest model of a TM, which contains the necessary ingredients to capture the physics we want to describe. 
The simplest such model is the following model of a lattice Dirac fermion
\beq
\label{eq:7}
H = t \gamma^0 \gamma^{\mu} \sin k_{\mu} + \Delta(\bk) \gamma^0, 
\eeq
where
\beq
\label{eq:8}
\Delta(\bk) = t (3 - \cos k_x - \cos k_y - \cos k_z), 
\eeq
and $\gamma^{\mu}$ are Dirac gamma matrices in, for example, the Weyl representation
\beq
\label{eq:9}
\gamma^0 = \tau^x, \,\, \gamma^i = - i \tau^y \sigma^i, \,\, i = 1,2,3. 
\eeq
This model describes two Weyl nodes of opposite chirality at the $\Gamma$-point in the BZ (the effects we will be discussing do not 
depend on the momentum-space separation between the Weyl nodes). 
Since a single Dirac point in the BZ is incompatible with the chiral symmetry, Eq.~\eqref{eq:7} also has an essential property, shared 
by all real Weyl and Dirac semimetals, that the chiral symmetry (chiral charge conservation) is only an approximate low-energy symmetry of Eq.~\eqref{eq:7}, which emerges when $H$ is expanded to linear order in $\bk$ near the $\Gamma$-point. 
In this case we have
\beq
\label{eq:10}
H = t \gamma^0 \gamma_{\mu} k_{\mu}, 
\eeq
and the chirality operator $\gamma^5 = i \gamma^0 \gamma^1 \gamma^2 \gamma^3 = \tau^z$ commutes 
with $H$, which is no longer true once nonlinear terms are included. 
This gives a finite (but small) chiral charge relaxation rate, which is an essential property of a Weyl or Dirac semimetal. 

We add a uniform magnetic field in the $z$-direction $\bB = B \hat z$, and choose the Landau gauge for the vector 
potential $\bA = x B \hat y$. 
We will ignore the Zeeman effect for simplicity. 
To find the eigenstates of $H$ in the presence of the magnetic field, we expand to first order in $k_{x,y}$, while keeping 
the full $k_z$ dependence. 
This is a good approximation in the regime of weak magnetic fields when $k_F \ell_B \gg 1$, which we will be interested in. 
For computational convenience we also make the following canonical transformation in the original Weyl representation 
of the gamma-matrices:
\beq
\label{eq:11}
\tau^{x,y} \rightarrow \sigma^z \tau^{x,y}, \,\, \sigma^{x,y} \rightarrow \tau^z \sigma^{x,y}. 
\eeq

This brings the Hamiltonian to the form
\beq
\label{eq:12}
H = t (\sigma^x \pi_x + \sigma^y \pi_y) + \hat m(k_z) \sigma^z, 
\eeq
where $\boldsymbol \pi = -i \bnabla + e \bA$ is the canonical momentum and 
\beq
\label{eq:13}
\hat m(k_z) = t \tau^z \sin k_z + \Delta (0,0,k_z) \tau^x. 
\eeq
Diagonalizing Eq.~\eqref{eq:12}, we find the eigenstate wavefunctions 
\beqa
\label{eq:14}
|n, s, p, k_y, k_z \rangle&=&z^{s p}_{n \uparrow \tau} (k_z) |n - 1, k_y, k_z, \upa, \tau \rangle \nonumber \\
&+&z^{s p}_{n \da \tau} (k_z) |n, k_y, k_z, \da, \tau \rangle, 
\eeqa
where $n$ is an integer Landau level index, $\upa, \da$ label the two eigenvalues of $\sigma^z$, $\tau = \pm$ are the two eigenvalues of
$\tau^z$, and $s, p = \pm$. 
Here and throughout sums over repeated indices will be implicit. 
The amplitudes $z^{s p}_{n \sigma \tau}(k_z)$ may be regarded as components of an eigenvector $|z^{s p}_n(k_z) \rangle = |v^{s p}_n(k_z) \rangle  \otimes |u^p(k_z) \rangle$, where 
\beqa
\label{eq:15}
&&|u^{p}_n(k_z) \rangle = \frac{1}{\sqrt{2}} \left(\sqrt{1 + p \frac{t \sin k_z}{m(k_z)}}, p \sqrt{1 - p \frac{t \sin k_z}{m(k_z)}} \right), \nonumber \\
&&|v^{s p}_n(k_z) \rangle = \frac{1}{\sqrt{2}} \left(\sqrt{1 + s p \frac{m(k_z)}{\epsilon_n(k_z)}}, s \sqrt{1 - s p \frac{m(k_z)}{\epsilon_n(k_z)}} \right). 
\nonumber \\
\eeqa
The corresponding energy eigenvalues are given by
\beq
\label{eq:16} 
\epsilon_{n s p}(k_z) = s \epsilon_n(k_z) = s \sqrt{2 \omega_B^2 n + m^2(k_z)}, 
\eeq
where $m(k_z) = 2 t |\sin k_z|$, and $\omega_B = t/ \ell_B$, for all $n \geq 1$. 
The lowest Landau level (LLL), corresponding to $n = 0$, is special: it does not have the $s$ label and its eigenenergy and the corresponding 
eigenvector are given by
\beq
\label{eq:17}
\epsilon_{0 p}(k_z) = - p m(k_z), 
\eeq
and 
\beq
\label{eq:18}
|v^p_0(k_z) \rangle = (0, 1).
\eeq

We add to the Hamiltonian Eq.~\eqref{eq:12} random impurity potential $V(\br)$, which we take to be of the Gaussian white noise form with 
$\langle V(\br) \rangle = 0$ and 
\beq
\label{eq:19}
\langle V(\br) V(\br') \rangle = \gamma^2 \delta(\br - \br'). 
\eeq
We take the impurity potential to be independent of the spin and orbital pseudospin indices. Physically this means that the impurities are 
taken to be nonmagnetic and the potential is smooth enough that its spatial variation on the scale of the unit cell of the crystal is negligible. 
The last assumption is not essential, but does simplify the subsequent calculations. 

We will evaluate the density response for the above model of a TM using the self-consistent Born approximation (SCBA) and 
the ladder approximation to perform the impurity averaging. 
This is a conserving approximation, meaning it preserves exact conservation laws and sum rules, and amounts physically to neglecting quantum interference effects. 
This is justified in the quasiclassical transport regime, which we will confine ourselves to: we assume that we are interested in the density response 
at length scales much longer than the inverse Fermi momentum and time scales much longer than the inverse Fermi energy; the impurity scattering is taken to be weak enough, so that $k_F \ell \gg 1$ and, as already mentioned, magnetic field is also assumed to be weak, which means 
$k_F \ell_B \gg 1$. Finally, we will assume that the Fermi energy is close to the Dirac point $\epsilon_F \ll t$ (but $\epsilon_F \tau \gg 1$), which defines the regime of a TM. 
The last condition ensures the near conservation of the chiral charge, as will be seen explicitly below. 

The calculation of the SCBA impurity self-energy in a similar model has already been discussed in detail in Ref.~\onlinecite{Burkov_lmr_prb}. 
We will thus omit the details of this calculation here and simply quote the result. 
One obtains that in the quasiclassical transport regime the impurity scattering rate is independent of both the Landau level index $n$ and 
the longitudinal momentum component $k_z$ and is given by the standard SCBA expression
\beq
\label{eq:20}
\frac{1}{\tau} = \frac{\pi \gamma^2 g}{2},
\eeq
where the density of states at the Fermi energy is given by
\beq
\label{eq:21}
g = \frac{\epsilon_F}{\pi t^2} \int_{-\pi}^{\pi} \frac{d k_z}{2 \pi} \Theta[\epsilon_F - m(k_z)], 
\eeq
$\Theta(x)$ being the Heaviside step function. 

We evaluate the density response function by summing the impurity ladder diagrams.
We start from the most general retarded density matrix response function, defined as
\beqa
\label{eq:22}
&&\chi_{\alpha_1 \alpha_2, \alpha_3 \alpha_4}(\br, t | \br', t') \nonumber \\
&=&- i \Theta(t - t') \langle [\varrho^\pdg_{\alpha_1 \alpha_2} (\br, t), \varrho^\dg_{\alpha_3 \alpha_4}(\br', t')] \rangle, 
\eeqa
where the density matrix is defined as
\beq
\label{eq:23}
\varrho_{\alpha_1 \alpha_2}(\br, t) = \Psi^\dg_{\alpha_2}(\br, t) \Psi^\pdg_{\alpha_1}(\br, t), 
\eeq
and $\alpha = (\sigma \tau)$ is a composite index, which encodes both the spin and orbital pseudospin labels. 

\begin{figure}[t]
\includegraphics[width=9cm]{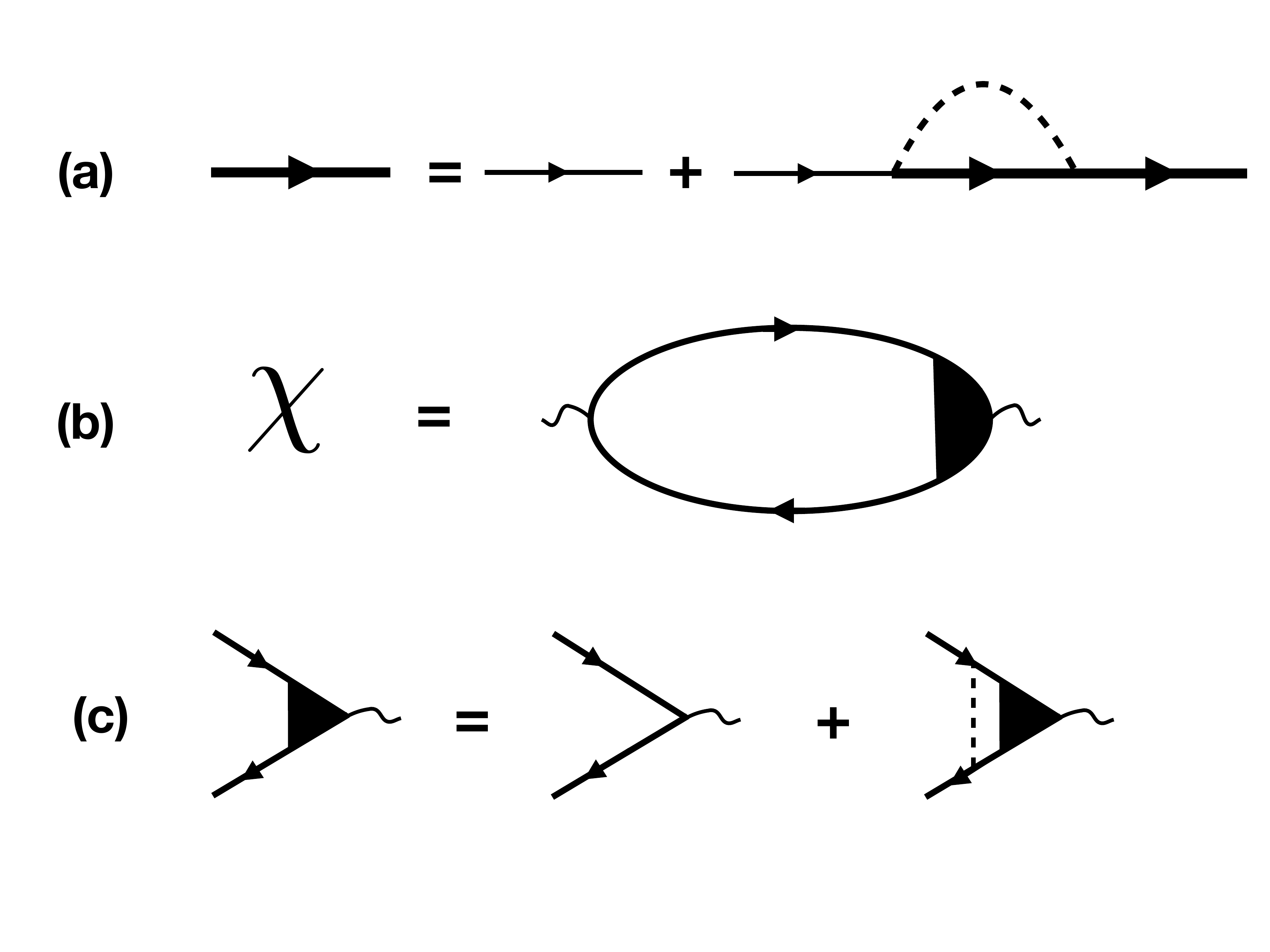}
\caption{Diagrammatic representation of (a) SCBA Green's function. Thin line represents the bare Green's function, thick line is the SCBA impurity-averaged Green's function and the dashed line represents the disorder potential correlator $\langle V(\br) V(\br') \rangle = \gamma^2 \delta(\br - \br')$. (b) Density response function $\chi$. (c) Bethe-Salpeter equation for the diffusion vertex $\cD$.}
\label{fig:1}
\end{figure}

The standard procedure to find the real-time response function Eq.~\eqref{eq:22} is to start from the corresponding imaginary-time 
response function
\beqa
\label{eq:24}
&&\chi_{\alpha_1 \alpha_2, \alpha_3 \alpha_4}(\br, \tau | \br', \tau') \nonumber \\
&=&- \cG_{\alpha_1 \alpha_3} (\br, \br', \tau - \tau') 
\cG_{\alpha_4 \alpha_2} (\br', \br, \tau' - \tau), 
\eeqa
where $\cG_{\alpha \alpha'}(\br, \br', \tau - \tau')$ is the exact imaginary-time Green's function, which depends on both $\br$ and $\br'$ separately 
due to both the lack of translational symmetry in the presence of a random impurity potential, and the lack of gauge invariance in the presence of 
an external magnetic field. 
One then performs impurity averaging, which restores translational invariance in the density response function and gives
\beq
\label{eq:25}
\chi_{\alpha_1 \alpha_2, \alpha_3 \alpha_4}(\bq , i \Omega) = \frac{1}{\beta} \sum_{i \omega} \cP_{\alpha_1 \alpha_2, \alpha_3 \alpha_4} (\bq, i \omega, 
i \omega  + i \Omega), 
\eeq
where 
\beqa
\label{eq:26}
&&\cP_{\alpha_1 \alpha_2, \alpha_3 \alpha_4}(\br - \br', i \omega, i \omega + i \Omega) \nonumber \\
&=&- \langle \cG_{\alpha_1 \alpha_3}(\br, \br' , i \omega + i \Omega) \cG_{\alpha_4 \alpha_2}(\br', \br, i \omega)\rangle,
\eeqa
is the impurity-averaged generalized polarization bubble, and $\beta = 1/T$ is the inverse temperature.
In the quasiclassical regime we are interested in, $\cP$ may be evaluated by summing all the SCBA diagrams for the impurity self-energy and 
the ladder vertex corrections, as shown in Fig.~\ref{fig:1}. 
The result of this diagram summation may be written in a shorthand matrix notation as
\beq
\label{eq:27}
\cP = \cP^0 \cD, 
\eeq
where $\cP^0$ is the bare polarization bubble, in which only the self-energy corrections are included
\beqa
\label{eq:28}
&&\cP^0_{\alpha_1 \alpha_2, \alpha_3 \alpha_4}(\br - \br', i \omega, i \omega + i \Omega) \nonumber \\
&=&-\cG_{\alpha_1 \alpha_3}(\br, \br', i\omega + i \Omega) \cG_{\alpha_4 \alpha_2}(\br', \br, i \omega).
\eeqa
$\cG_{\alpha \alpha'}(\br, \br', i\omega)$ here is the disorder-averaged SCBA Green's function, which still depends 
on $\br$ and $\br'$ separately since it is a gauge-dependent quantity in the presence of an external magnetic field. 
The vertex part $\cD$, which is also known as the diffusion propagator, or diffuson, satisfies the following Bethe-Salpeter
equation
\beq
\label{eq:29}
\cD = 1 + \cI \cD, 
\eeq
where $\cI \equiv \gamma^2 \cP^0$. 
The solution of this equation is 
\beq
\label{eq:30}
\cD = (1 - \cI)^{-1}. 
\eeq
To obtain the real-time retarded response function we analytically continue to real frequency $i \Omega \rightarrow \Omega + i \eta$, which gives
\beqa
\label{eq:31}
&&\chi(\bq, \Omega) = \int_{-\infty}^{\infty} \frac{d \epsilon}{2 \pi i} n_F(\epsilon) 
\left[\cP(\bq, \epsilon + i \eta, \epsilon + \Omega + i \eta) \right. \nonumber \\
&-&\left.\cP(\bq, \epsilon - i \eta, \epsilon + \Omega + i \eta) + \cP(\bq, \epsilon - \Omega - i \eta, \epsilon + i \eta) \right. \nonumber \\
&-&\left.\cP(\bq, \epsilon - \Omega - i \eta, \epsilon - i \eta)\right]. 
\eeqa
In the low-frequency limit, when $\Omega \ll \epsilon_F$, this simplifies to
\beqa
\label{eq:32}
\chi(\bq, \Omega)&=&- \frac{i \Omega}{2 \pi} \cP(\bq, - i \eta, \Omega + i \eta) \nonumber \\
&-&\frac{1}{\pi} \int_{-\infty}^{\infty} d \epsilon \, n_F(\epsilon) \textrm{Im} \cP(\bq, \epsilon + i \eta, \epsilon + \Omega + i \eta)  \nonumber \\
&\equiv&\chi^I(\bq, \Omega) + \chi^{II}(\bq, \Omega). 
\eeqa
The physical meaning of the two contributions to the density response function, $\chi^I$ and $\chi^{II}$, is that $\chi^I$ arises from states on the 
Fermi surface, while all filled states contribute to $\chi^{II}$. 
$\chi^{II}$ thus represents equilibrium part of the response and is easily shown to be a diagonal matrix, with the nonzero matrix elements
equal to $-g$. 
On the other hand, $\chi^I$ represents the dynamical nonequilibrium part of the density response and is given by
\beq
\label{eq:33}
\chi^I(\bq, \Omega) = - \frac{i \Omega}{2 \pi \gamma^2} \cI(\bq, \Omega) \cD(\bq, \Omega), 
\eeq
where
\beqa
\label{eq:34}
&&\cI_{\alpha_1 \alpha_2, \alpha_3 \alpha_4}(\bq, \Omega) = \frac{\gamma^2}{L^3} \int d^3 r \, d^3 r' e^{- i \bq \cdot (\br - \br')} \nonumber \\
&\times&G^R_{\alpha_1 \alpha_3}(\br, \br', \Omega) G^A_{\alpha_4 \alpha_2}(\br', \br, 0), 
\eeqa
$G^{R, A}$ being the retarded and advanced real-time impurity-averaged SCBA Green's functions. They are explicitly given by
\beqa
\label{eq:35}
&&G^{R,A}_{\alpha \alpha'}(\br, \br', \omega) \nonumber \\
&=&\sum_{n s p k_y k_z} \frac{\langle \br, \alpha | n, s, p, k_y, k_z \rangle \langle n, s, p, k_y, k_z | \br', \alpha'
\rangle}{\omega - \xi_{n s p}(k_z) \pm i/2 \tau}, \nonumber \\
\eeqa
where $\xi_{n s p}(k_z) \equiv \epsilon_{n s p}(k_z) - \epsilon_F$. 

For a general direction of the wavevector $\bq$, the evaluation of $\cI(\bq, \Omega)$ is severely complicated by the fact that contributions of 
different Landau levels are mixed in Eq.~\eqref{eq:34}. 
This is not the case only when $\bq = q \hat z$, when translational symmetry in the $xy$-plane leads to decoupling of the individual Landau level contributions. Fortunately, this is in fact the case of primary interest to us, since the chiral anomaly leads to unusual transport phenomena in the 
direction of the magnetic field. 
Thus we will take $\bq = q \hat z$ henceforth. 

In this case the evaluation of $\cI(\bq, \Omega)$ is relatively straightforward, particularly in the weak magnetic field regime $k_F \ell_B \gg 1$ that we 
are interested in. 
An additional simplification arises from the fact that we are not interested in the whole $16 \times 16$ matrix $\cI$, which contains a lot of unnecessary information. 
We are interested only in the response of conserved, or nearly conserved, quantities, which will always dominate everything else at long times 
and long distances. 
In a generic TM, we expect only two such quantities to exist: the electric charge, which is strictly conserved, and the chiral charge, whose near conservation is a defining property of a TM, as discussed above. 
Thus we may project the original $16 \times 16$ matrix onto the $2 \times 2$ subspace, describing the coupled transport of the electric and the chiral 
charge, which is accomplished as
\beq
\label{eq:36}
\cI_{a b}(\bq, \Omega) = \frac{1}{4} \Gamma^a_{\alpha_2 \alpha_1} \cI_{\alpha_1 \alpha_2, \alpha_3 \alpha_4}(\bq, \Omega) 
\Gamma^b_{\alpha_3 \alpha_4}, 
\eeq
where $a, b = 0, 5$, corresponding to the electric $(0)$ or chiral $(5)$ charges, and $\Gamma^{a,b}$ are the corresponding operators, i.e.
\beq
\label{eq:37}
\Gamma^0 = \tau^0 \sigma^0 = 1,\,\, \Gamma^5 = \tau^z \sigma^0 = \tau^z. 
\eeq
After a tedious, but straightforward, calculation, we obtain
\begin{widetext}
\beqa
\label{eq:38}
\cI_{00}(q, \Omega)&=&\frac{i}{2 q \ell} \ln \left(\frac{1 - i \Omega \tau - i q \ell}{1 - i \Omega \tau + i q \ell} \right), \nonumber \\
\cI_{55}(q, \Omega)&=&\left[\frac{i}{2 q \ell} - \frac{i (\epsilon_F/t)^2 (1 - i \Omega \tau)^4}{8 (q \ell)^5} \right] \ln \left(\frac{1 - i \Omega \tau - i q \ell}{1 - i \Omega \tau + i q \ell} \right) - \frac{(\epsilon_F/t)^2 (1 - i \Omega \tau)}{12 (q \ell)^2} + \frac{(\epsilon_F / t)^2 (1 - i \Omega \tau)^3}{4 (q \ell)^4}, \nonumber \\
\cI_{05}(q, \Omega)&=&\cI_{50}(q, \Omega) = \frac{i}{2 (k_F \ell_B)^2} \frac{q \ell}{(1 - i \Omega \tau)^2 + (q \ell)^2}.
\eeqa
\end{widetext}
Substituting this into Eq.~\eqref{eq:33}, we obtain the dynamical nonequilibrium contribution to the density response $\chi^I(q, \Omega)$, while
the equilibrium contribution is a diagonal matrix given by
\beq
\label{eq:39}
\chi^{II}_{00}(q, \Omega) = \chi^{II}_{55}(q, \Omega) = - g, 
\eeq
as already mentioned above. 

A comment is in order here. As can be seen from Eq.~\eqref{eq:38}, only the off-diagonal matrix element $\cI_{05}$ depends on the magnetic field. 
This is true in the quasiclassical limit $k_F \ell_B \gg 1$ only, and is a consequence of the fact that in this limit we may ignore the effect of the magnetic 
field on the density of states. 
Summation over the Landau level index $n$, which arises when evaluating Eq.~\eqref{eq:38}, may in this case be replaced by integration and the magnetic field dependence disappears to leading order in $1/k_F \ell_B$. 
In contrast, the off-diagonal matrix element $\cI_{05}$ arises entirely from the contribution of the $n = 0$ Landau level. 
This contribution is proportional to $1/(k_F \ell_B)^2$, but leads to large effects at long length scales and long times, as will be seen below, provided 
$\tau_c/ \tau \gg 1$. 

Eqs.~\eqref{eq:32}, \eqref{eq:33}, \eqref{eq:38} and \eqref{eq:39} give a general expression for the density response function of a TM in the quasiclassical regime
\beq
\label{eq:39.5}
\chi(q, \Omega) = -g [ i \Omega \tau \cI(q, \Omega) \cD(q, \Omega) + 1]. 
\eeq
This expression is valid in either diffusive $\Omega \tau, q \ell \ll 1$ or ballistic $\Omega \tau, q \ell \gg 1$ limits and may be used, in particular, 
to study the ballistic-diffusive crossover regime. 
We will start by analyzing the two limits. 
\subsection{Ballistic regime}
\label{sec:2a}
In this regime all components of the matrix $\cI$ are small and thus $\cD \approx 1$. 
Physically this means that we are looking at short length and time scales at which the impurity scattering may be ignored. 
While the response function $\chi(q, \Omega)$ is a $2 \times 2$ matrix, only its $\chi_{00}(q, \Omega)$ component describes observable density response. 
Taking the limit $\Omega \tau, q \ell \rightarrow \infty$ in Eqs.~\eqref{eq:38}, \eqref{eq:39.5} we obtain
\beq
\label{eq:40}
\chi_{00}(q, \Omega) = -g \left[1 + \frac{\Omega}{2 q t} \ln \left(\frac{\Omega - q t + i \eta}{\Omega + q t + i \eta} \right) \right]. 
\eeq
This is just the familiar Lindhard function (in the limit $q \ll k_F$ and $\Omega \ll \epsilon_F$), describing the density response of a clean Fermi liquid with the Fermi velocity $t$. 
The imaginary part of $\chi_{00}(q, \Omega)$, which is determined by the branch cuts of the Lindhard function
\beq
\label{eq:41}
\textrm{Im} \chi_{00}(q, \Omega) = g \frac{\pi \Omega}{2 q t} \,\Theta(q t - |\Omega|), 
\eeq
describes the excitation spectrum of the Fermi liquid, which forms a particle-hole continuum. 
Thus in the ballistic regime and in the weak magnetic field limit chiral anomaly has no effect on the density response of a TM [its effects appear only at order $1/(k_F \ell_B)^2$, which is negligible compared to Eq.~\eqref{eq:40}]. 
This of course will no longer be true if we tune the Fermi energy to zero (i.e. to the ideal Weyl or Dirac semimetal limit), but this is a fine-tuned, 
non-generic situation, and is of somewhat less interest for this reason. 

\subsection{Diffusive regime}
\label{sec:2b}
The situation is much more interesting in the diffusive limit $\Omega \tau, q \ell \ll 1$. 
In this case $\cI \approx 1$, and multiple impurity scattering needs to be taken into account. 
Expanding in Taylor series in $\Omega \tau$ and $q \ell$, we obtain
\beqa
\label{eq:42}
\cI_{00}(q, \Omega)&\approx&1 + i \Omega \tau - D q^2 \tau, \nonumber \\
\cI_{05}(q, \Omega)&=&\cI_{50}(q, \Omega) \approx i \Gamma q \tau, \nonumber \\
\cI_{55}(q, \Omega)&\approx&1 + i \Omega \tau - \tau/ \tau_c - D q^2 \tau. 
\eeqa
Here $D = t^2 \tau/3 = t \ell/ 3$ is the diffusion constant, 
\beq
\label{eq:43}
\Gamma = \frac{e B}{2 \pi^2 g} = \frac{t}{2 (k_F \ell_B)^2}, 
\eeq
is a new transport coefficient, which describes the chiral-anomaly-induced coupling between the electric and the chiral charge 
densities, and 
\beq
\label{eq:44}
\frac{1}{\tau_c} = \frac{\epsilon_F^2}{20 \, t^2 \tau}, 
\eeq
is the chiral charge relaxation rate.
Note that the fact the chiral charge relaxation rate vanishes in the limit $\epsilon_F \rightarrow 0$ is a consequence of our assumption that the 
impurity potential is diagonal in the spin and orbital indices and thus commutes with the chiral charge operator $\gamma^5 = \tau^z$. 
In general this is not the case and we can expect some residual chiral charge relaxation even in the $\epsilon_F \rightarrow 0$ limit. 

In the diffusive regime the dynamics of the density response is determined by the poles of the diffusion propagator $\cD$, instead of the 
branch cuts of the response function, as in the ballistic limit. 
From Eq.~\eqref{eq:42}, the inverse diffusion propagator is given by
\beqa
\label{eq:45}
\cD^{-1}(q, \Omega) = \left(
\begin{array}{cc}
-i \Omega \tau + D q^2 \tau & - i \Gamma q \tau \\
-i \Gamma q \tau & -i \Omega \tau + \tau/ \tau_c + D q^2 \tau
\end{array}
\right), \nonumber \\
\eeqa
The zeros of the determinant of this matrix determine the eigenmode frequencies
\beq
\label{eq:46}
\Omega_{\pm} = \pm \Omega_0 - i (D q^2 + 1/2 \tau_c), 
\eeq
where 
\beq
\label{eq:47}
\Omega_0 = \sqrt{\Gamma^2 q^2 - 1/4 \tau_c^2}. 
\eeq
The diffusion propagator itself may then be written as 
\beqa
\label{eq:48}
\cD(q, \Omega)&=&\frac{1/\tau}{(\Omega - \Omega_+)(\Omega - \Omega_-)} \nonumber \\
&\times&\left(
\begin{array}{cc}
i \Omega - D q^2 - 1/\tau_c & - i \Gamma q \\
- i \Gamma q & i \Omega - D q^2 
\end{array}
\right). 
\eeqa

Taking into account that in the diffusive regime $\cI \approx 1$, we obtain from Eq.~\eqref{eq:39.5}
\beq
\label{eq:49}
\chi(q, \Omega) \approx -g [ i \Omega \tau \cD(q, \Omega) + 1], 
\eeq
which gives the following explicit expression for the $00$ component of the matrix response function, which corresponds to the observable 
electric charge density response 
\beq
\label{eq:50}
\chi_{00}(q, \Omega) = g \left[\frac{\Omega (\Omega + i/ \tau_c + i D q^2)}{(\Omega - \Omega_+)(\Omega - \Omega_-)} - 1\right]. 
\eeq

We now note that the frequency $\Omega_0$ is purely imaginary at the smallest momenta when 
\beq
\label{eq:51}
q < \frac{1}{2 \Gamma \tau_c} = \frac{L_a}{2 L_c^2} \equiv \frac{1}{L_*}, 
\eeq
where we have introduced two new length scales 
\beq
\label{eq:52}
L_c = \sqrt{D \tau_c},
\eeq
which has the meaning of the chiral charge diffusion length and 
\beq
\label{eq:53} 
L_a = \frac{D}{\Gamma} = \frac{2}{3} \ell (k_F \ell_B)^2. 
\eeq
$L_a$ is a magnetic-field-related length scale, distinct from the magnetic length, which arises from the chiral anomaly. 
It is a long hydrodynamic length scale in the weak magnetic field regime, in the sense that $L_a \gg \ell$, but it may still be 
much smaller that either the chiral charge diffusion length $L_c$ or the sample size $L$. 
In fact, the ratio $L_c/L_a$ quantifies the strength of the chiral-anomaly-related density response phenomena, as will be seen below. 

Thus when $q < 1/L_*$ the eigenfrequencies of the diffusion propagator are purely imaginary, which corresponds to ordinary diffusion (nonpropagating) 
modes. 
However, when $q > 1/L_*$ (which may be a very small momentum when the ratio $L_c/L_a$ is large), $\Omega_0$ is real, which signals the emergence of a pair of propagating modes in this regime. 
The modes are only weakly damped as long as 
\beq
\label{eq:54}
\Omega_0 \approx \Gamma q > D q^2, 
\eeq
which defines the upper limit on the wavevector $q = 1/L_a$, above which the propagating modes disappear. 
The propagating modes thus exist in the interval 
\beq
\label{eq:55}
1/L_* < q < 1/L_a. 
\eeq
This interval is significant when $L_c/ L_a \gg 1$. 

Within this interval of $q$ the density response function takes the following approximate form 
\beq
\label{eq:56}
\chi_{00}(q, \Omega) = g \frac{\Omega_0^2}{(\Omega + i D q^2)^2 - \Omega_0^2}, 
\eeq
where $\Omega_0 = \Gamma q$. 
This is the density response function of an effective 1D system with the Fermi velocity $\Gamma = t/2 (k_F \ell_B)^2 \ll t$. 
Note that this is very different from the 1D response one would obtain in a TM in the quantum limit $k_F \ell_B < 1$, 
when only the lowest $n = 0$ Landau level contributes to the density response. 
In this case one gets $N_{\phi}$ 1D modes, which correspond to $N_{\phi}$ orbital states within the LLL. 
The Fermi velocity of these 1D modes is equal to the microscopic Fermi velocity $t$. 
In our case, while the ultimate origin of the 1D dynamics is still the LLL, its emergence is only possible in the diffusive regime and 
thus requires multiple impurity scattering. 
The corresponding Fermi velocity $\Gamma$ is proportional to the applied magnetic field and is much smaller than $t$ in the quasiclassical 
weak-field regime. 
Such effectively 1D density response, with propagating rather than diffusive density dynamics, which exists in a 3D metal with a large Fermi surface 
($k_F \ell \gg 1$) in a weak magnetic field ($k_F \ell_B \gg 1$) is a truly unique feature of TM and should be regarded as their true smoking-gun characteristic. 

On the other hand, when $L_a > L_c$, propagating modes do not exist for any $q$ and one obtains a pair of standard diffusion modes
\beq
\label{eq:57}
\Omega_+ = - i D q^2, \,\, \Omega_- = - i D q^2 - i/\tau_c, 
\eeq
which correspond to independent diffusion of the electric and the chiral charge densities. 

\section{Transport in topological metals}
\label{sec:3}
It is very useful to also look at the transport properties, which follow from the density response, described in Section~\ref{sec:2}. 
In addition to providing further insight into the physical meaning of the results, discussed in the previous section, this will also 
allow us to calculate experimentally measurable physical quantities, such as the frequency- and scale-dependent conductivity. 
\subsection{Scale-dependent conductance}
\label{sec:3a}
It is easy to show that Eqs.~\eqref{eq:48} and \eqref{eq:49} for the diffusion propagator and the generalized density response function are equivalent 
to the following transport equation in real space and time, that the electric $n_0$ and chiral $n_5$ charge densities must satisfy
\beqa
\label{eq:58}
\frac{\partial n_0}{\partial t}&=&D \bnabla^2 (n_0 + g V_0) + \bGamma \cdot \bnabla (n_5 + g V_5), \nonumber \\
\frac{\partial n_5}{\partial t}&=&D \bnabla^2 (n_5 + g V_5) - \frac{n_5 + g V_5}{\tau_c} + \bGamma \cdot \bnabla (n_0 + g V_0), \nonumber \\
\eeqa
where $V_0$ and $V_5$ are external electric and chiral potentials correspondingly and we have generalized to an arbitrary magnetic field 
direction, which is why the coefficient $\bGamma \propto \bB$ has become a vector. 
The chiral potential $V_5$ may arise, for example, in a situation when the inversion symmetry is broken, in which case the Weyl nodes of different chirality will generally be located at different energies, $V_5$ being precisely this energy difference. 
Otherwise this should simply be regarded as a fictitious potential, which couples linearly to the chiral charge $n_c$. 
Indeed, Fourier transforming Eq.~\eqref{eq:58} we obtain
\beqa
\label{eq:59}
\cD^{-1}(q, \Omega) \left(
\begin{array}{c}
n_0 \\ n_5
\end{array}
\right) = -g \left[ i \Omega \tau + \cD^{-1}(q, \Omega) \right] \left(
\begin{array}{c}
V_0 \\ V_5
\end{array}
\right). \nonumber \\
\eeqa
This gives
\beqa
\label{eq:60}
\left(
\begin{array}{c}
n_0 \\ n_5
\end{array}
\right) = - g [ i \Omega \tau \cD(q, \Omega) + 1] \left(
\begin{array}{c}
V_0 \\ V_5
\end{array}
\right), 
\eeqa
which is equivalent to Eq.~\eqref{eq:49}. 

Solving Eq.~\eqref{eq:58} in the steady state, assuming a uniform sample of linear size $L$, attached to normal metal leads (in which the chiral electrochemical potential $n_5 + g V_5 = 0$) in the $z$-direction (i.e. the current flows along the magnetic field), one obtains the following 
expression for the scale-dependent sample conductance~\cite{Altland15,Burkov_gphe}
\beq
\label{eq:61}
G(L) = \frac{e^2 N_{\phi}}{2 \pi} F(L/L_a, L/L_c), 
\eeq
where the scaling function $F(x,y)$ is given by
\beq
\label{eq:62}
F(x,y) = \frac{(1 + y^2/x^2)^{3/2}}{\frac{y^2}{2 x} \sqrt{1 + y^2/x^2} + \tanh \left(\frac{x}{2} \sqrt{1 + y^2/x^2}\right)}. 
\eeq
This scaling function exhibits crossover behaviors which exactly match the corresponding crossovers in the wavevector dependence 
of the diffusion modes, described in Section~\ref{sec:2}. 

Indeed, when $x \ll y$, which means $L_a \gg L_c$, we have $F(x, y) \approx 2/x$, which gives 
\beq
\label{eq:63}
G(L) \approx e^2 g D L = \sigma L, 
\eeq
which is simply the standard Ohmic conductance, with a small magnetic-field dependent correction, which goes as $(L_c/L_a)^2$, and which we have ignored here for the sake of brevity.~\cite{Burkov_gphe}
This corresponds to the regime, in which we have two independent diffusion modes, given by Eq.~\eqref{eq:57}, corresponding to independent diffusion of the electric and the chiral charges. 

On the other hand, when $L_a \ll L_c$, or $x \gg y$, we obtain 
\beq
\label{eq:64}
F(x, y) \approx \frac{1}{y^2/ 2 x + \tanh(x/2)}.
\eeq
This exhibits a regime of quasiballistic conductance with 
\beq
\label{eq:65}
G(L) \approx \frac{e^2 N_{\phi}}{2 \pi}, 
\eeq
which is realized when 
\beq
\label{eq:66}
L_a < L < L_*.
\eeq
This corresponds precisely to the range of the wavevectors $q$ in Eq.~\eqref{eq:55}, for which propagating modes exist when $L_a \ll L_c$. 
Thus, one of the observable manifestations of the existence of quasi-1D propagating modes in a TM is the quasiballistic 
conductance, given by Eq.~\eqref{eq:65}. 

It is instructive to see what the quasiballistic conductance regime corresponds to directly in terms of the transport equations Eq.~\eqref{eq:58}. 
In this regime both the second derivative $D \bnabla^2 n_{0,5}$ and the relaxation $n_5/\tau_c$ terms may be ignored and we obtain
\beqa
\label{eq:67}
\frac{\partial n_0}{\partial t}&=&\Gamma \frac{\partial n_5}{\partial z}, \nonumber \\
\frac{\partial n_5}{\partial t}&=&\Gamma \frac{\partial n_0}{\partial z}. 
\eeqa
Introducing the left- and right-handed charges as $n_{R,L} = (n_0 \pm n_5)/2$ we obtain 
\beqa
\label{eq:68}
\frac{\partial n_R}{\partial t}&=&\Gamma \frac{\partial n_R}{\partial z}, \nonumber \\
\frac{\partial n_L}{\partial t}&=&-\Gamma \frac{\partial n_L}{\partial z}. 
\eeqa
Eq.~\eqref{eq:68} describes two chiral bosonic density modes, which propagate along and opposite to the direction of the applied magnetic field. 
Such ``bosonization" of the electron dynamics, which occurs in a 3D metal in a weak quasiclassical magnetic field, is a characteristic smoking-gun feature of a TM. 

Eq.~\eqref{eq:68} means, in particular, that a density disturbance, created in a TM in magnetic field, with split into two chiral modes, which 
will propagate ballistically in opposite directions, spatially separating electrons of different chirality. 
It might be possible to detect this effect optically.~\cite{Gedik17}
\subsection{Optical conductivity}
\label{sec:3b}
Optical conductivity of TM has been studied before, with a focus mostly on the interband transition effects.~\cite{Carbotte_optical,Mele_optical,Roy_optical,Sun_optical,Felser_optical}
Here we will demonstrate that low-frequency intraband optical conductivity is qualitatively affected by the chiral anomaly, which has not been noticed before. 

\begin{figure}[t]
\includegraphics[width=9cm]{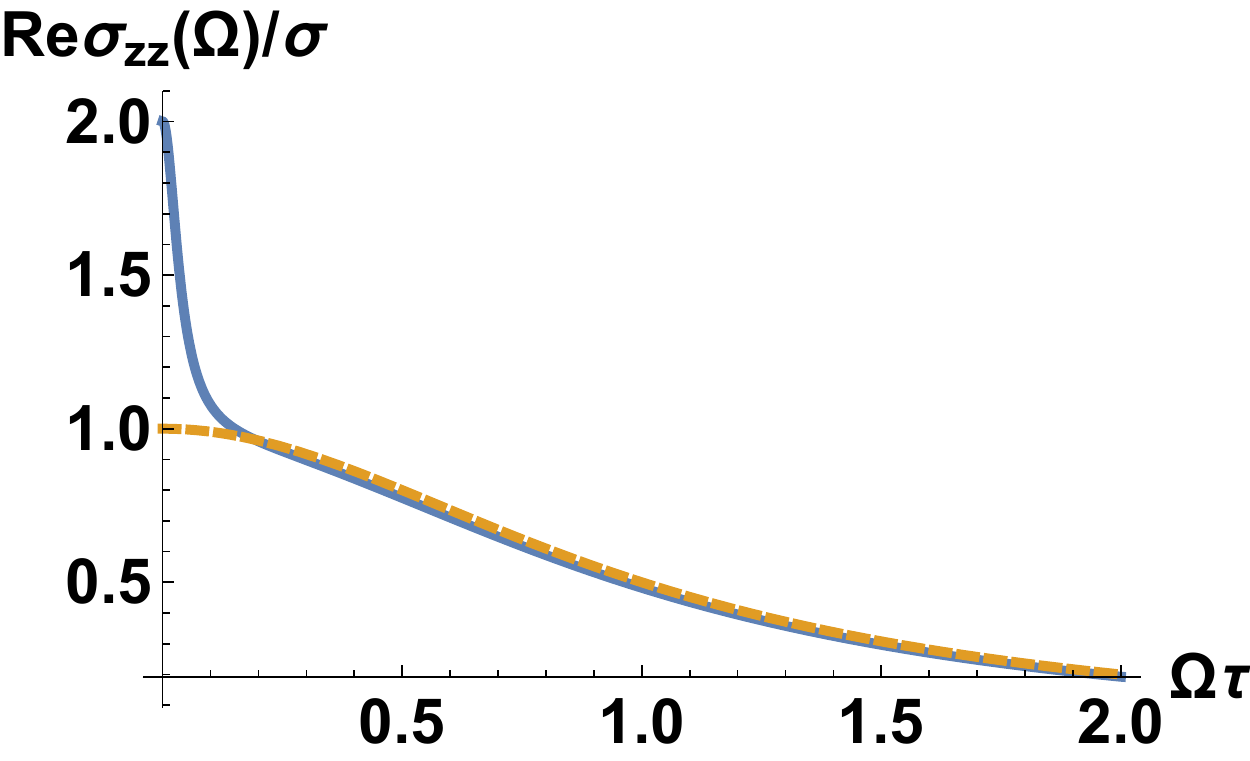}
\caption{(Color online) Frequency-dependent conductivity for $L_c/L_a = 1$ (solid line) and $L_c/L_a = 0$ (dashed line), and $\tau/\tau_c = 0.04$.}
\label{fig:2}
\end{figure}

From the general expression for the density response function Eq.~\eqref{eq:39.5} we may easily obtain the frequency-dependent conductivity. 
Indeed, electric charge conservation requires that 
\beq
\label{eq:69}
\sigma_{zz}(\Omega) = - e^2 \lim_{q \rightarrow 0} \frac{i \Omega}{q^2} \chi_{00}(q, \Omega). 
\eeq
A straightforward calculation then gives
\beq
\label{eq:70}
\sigma_{zz}(\Omega) = \frac{\sigma}{1 - i \Omega \tau} \frac{1 - i \Omega \tau_c + (L_c/ L_a)^2}{1 - i \Omega \tau_c},
\eeq
where $\sigma = e^2 g D$ is the zero-field DC conductivity. 
Evaluating the real part, one obtains
\beq
\label{eq:71}
\textrm{Re} \,\sigma_{zz}(\Omega) = \frac{\sigma}{1 + \Omega^2 \tau^2} \left[1 + \left(\frac{L_c}{L_a}\right)^2 \frac{1 - \Omega^2 \tau \tau_c}{1 + \Omega^2 \tau_c^2} \right]. 
\eeq
Eq.~\eqref{eq:71} is one of the main new results of this paper. 
The prefactor in Eq.~\eqref{eq:71} is the standard Drude expression for the optical conductivity of a metal. 
The part in the square brackets is a correction that arises in a TM as a consequence of the chiral anomaly. 
This correction represents transfer of the spectral weight from high frequencies into a new low-frequency peak, whose width scales with the 
chiral charge relaxation rate $1/\tau_c$, while height is proportional to the ratio $(L_c/L_a)^2$. 
Importantly, Eq.~\eqref{eq:71} satisfies the exact $f$-sum rule
\beq
\label{eq:72}
\int_0^{\infty} d \Omega\,\, \textrm{Re}\, \sigma_{zz}(\Omega) = \frac{\pi \sigma}{2 \tau}, 
\eeq
which means that the appearance of the new low-frequency peak indeed represents spectral weight transfer, as it should, see Fig.~\ref{fig:2}. 

It is instructive to examine the high-frequency limit of Eq.~\eqref{eq:71}, namely when $\Omega > 1/ \tau_c, 1/\sqrt{\tau \tau_c}$. 
In this limit we obtain
\beq
\label{eq:73}
\textrm{Re}\, \sigma_{zz}(\Omega) \approx \frac{\sigma}{1 + \Omega^2 \tau^2} \left[1 - \frac{1}{3} \left(\frac{\ell}{L_a} \right)^2\right]. 
\eeq
The negative second term in the square brackets expresses the reduction of the spectral weight at high frequencies, induced by the chiral anomaly.
We note that while formally the whole expression may become negative for $L_a \ll \ell$, this would be outside of the regime of validity of our theory, which assumes weak magnetic field regime $k_F \ell_B \gg 1$ and thus $L_a \gg \ell$. Within this regime, the real part of the optical conductivity is 
always positive, as it should be.
\section{Discussion and conclusions}
\label{sec:4}
In this paper we have studied density response in TM and the corresponding experimentally observable phenomena. 
We have argued that one of the truly unique features of TM is the existence of propagating density modes, which are induced 
by the combined effect of the chiral anomaly and impurity scattering. 
The modes exist only in the diffusive limit and disappear in the ballistic regime. 
We have demonstrated that one of the observable manifestations of the existence of such propagating modes is the highly nontrivial 
scaling of the conductance of a TM with the sample size, first pointed out by Altland and Bagrets.~\cite{Altland15}
We have also demonstrated an entirely new phenomenon, namely a nontrivial frequency dependence of the optical conductivity, which exhibits 
transfer of the spectral weight from high frequencies, greater than $1/\sqrt{\tau \tau_c}$, into a new non-Drude low-frequency peak of width $1/\tau_c$. 
The existence of this new narrow peak in the optical conductivity is a smoking-gun consequence of the chiral anomaly in TM. 

One issue we have not touched upon in this paper is the effect of the electron-electron, in particular long-range Coulomb, interactions. 
One might worry that the Coulomb interactions could push the linearly-dispersing sound-like mode Eq.~\eqref{eq:54} to the 
plasma frequency, as happens in the case of the ordinary electronic zero sound mode, if short-range interactions are replaced by Coulomb interactions. 
This does not happen in our case, however, since the existence of the sound-like mode has nothing to do with the electron-electron interactions. 
Its physical origin lies in the effective ``one-dimensionalization" of the electron dynamics in a dirty TM in the presence of even a weak magnetic field. What this means is that the LLL dominates the density response at long times and long distances even when many higher Landau levels are occupied since the dynamics is ballistic in the LLL while it is diffusive in the higher Landau levels. 
This picture has nothing to do with the electron-electron interactions and will not be significantly modified by them, just as the ordinary low-energy particle-hole continuum in a clean Fermi liquid is not significantly affected by the interactions. 
The frequency of the plasmon modes $\Omega_P \sim \sqrt{ e^2 t^2 g}$ is not significantly affected by a weak applied magnetic 
field~\cite{Panfilov14,Xiao15} and is much larger than the frequency of the low-energy chiral density mode $\Omega_0 = \Gamma q$, which arises within the low-energy particle-hole continuum of the clean metal. This means that the two modes do not interact with each other in any significant way. 
However, the issue of collective plasmon modes in a dirty TM is interesting in its own right and will be addressed in a future 
publication. 
\begin{acknowledgments}
We thank Xi Dai for a useful discussion. Financial support was provided by Natural Sciences and Engineering Research Council (NSERC) of Canada. 
\end{acknowledgments}
\bibliography{references}

\begin{thebibliography}{44}%
\makeatletter
\providecommand \@ifxundefined [1]{%
 \@ifx{#1\undefined}
}%
\providecommand \@ifnum [1]{%
 \ifnum #1\expandafter \@firstoftwo
 \else \expandafter \@secondoftwo
 \fi
}%
\providecommand \@ifx [1]{%
 \ifx #1\expandafter \@firstoftwo
 \else \expandafter \@secondoftwo
 \fi
}%
\providecommand \natexlab [1]{#1}%
\providecommand \enquote  [1]{``#1''}%
\providecommand \bibnamefont  [1]{#1}%
\providecommand \bibfnamefont [1]{#1}%
\providecommand \citenamefont [1]{#1}%
\providecommand \href@noop [0]{\@secondoftwo}%
\providecommand \href [0]{\begingroup \@sanitize@url \@href}%
\providecommand \@href[1]{\@@startlink{#1}\@@href}%
\providecommand \@@href[1]{\endgroup#1\@@endlink}%
\providecommand \@sanitize@url [0]{\catcode `\\12\catcode `\$12\catcode
  `\&12\catcode `\#12\catcode `\^12\catcode `\_12\catcode `\%12\relax}%
\providecommand \@@startlink[1]{}%
\providecommand \@@endlink[0]{}%
\providecommand \url  [0]{\begingroup\@sanitize@url \@url }%
\providecommand \@url [1]{\endgroup\@href {#1}{\urlprefix }}%
\providecommand \urlprefix  [0]{URL }%
\providecommand \Eprint [0]{\href }%
\providecommand \doibase [0]{http://dx.doi.org/}%
\providecommand \selectlanguage [0]{\@gobble}%
\providecommand \bibinfo  [0]{\@secondoftwo}%
\providecommand \bibfield  [0]{\@secondoftwo}%
\providecommand \translation [1]{[#1]}%
\providecommand \BibitemOpen [0]{}%
\providecommand \bibitemStop [0]{}%
\providecommand \bibitemNoStop [0]{.\EOS\space}%
\providecommand \EOS [0]{\spacefactor3000\relax}%
\providecommand \BibitemShut  [1]{\csname bibitem#1\endcsname}%
\let\auto@bib@innerbib\@empty
\bibitem [{\citenamefont {Armitage}\ \emph {et~al.}(2018)\citenamefont
  {Armitage}, \citenamefont {Mele},\ and\ \citenamefont
  {Vishwanath}}]{Weyl_RMP}%
  \BibitemOpen
  \bibfield  {author} {\bibinfo {author} {\bibfnamefont {N.~P.}\ \bibnamefont
  {Armitage}}, \bibinfo {author} {\bibfnamefont {E.~J.}\ \bibnamefont {Mele}},
  \ and\ \bibinfo {author} {\bibfnamefont {A.}~\bibnamefont {Vishwanath}},\
  }\href {\doibase 10.1103/RevModPhys.90.015001} {\bibfield  {journal}
  {\bibinfo  {journal} {Rev. Mod. Phys.}\ }\textbf {\bibinfo {volume} {90}},\
  \bibinfo {pages} {015001} (\bibinfo {year} {2018})}\BibitemShut {NoStop}%
\bibitem [{\citenamefont {Hasan}\ \emph {et~al.}(2017)\citenamefont {Hasan},
  \citenamefont {Xu}, \citenamefont {Belopolski},\ and\ \citenamefont
  {Huang}}]{Hasan_ARCMP}%
  \BibitemOpen
  \bibfield  {author} {\bibinfo {author} {\bibfnamefont {M.~Z.}\ \bibnamefont
  {Hasan}}, \bibinfo {author} {\bibfnamefont {S.-Y.}\ \bibnamefont {Xu}},
  \bibinfo {author} {\bibfnamefont {I.}~\bibnamefont {Belopolski}}, \ and\
  \bibinfo {author} {\bibfnamefont {S.-M.}\ \bibnamefont {Huang}},\ }\href@noop
  {} {\bibfield  {journal} {\bibinfo  {journal} {Annual Review of Condensed
  Matter Physics}\ }\textbf {\bibinfo {volume} {8}} (\bibinfo {year}
  {2017})}\BibitemShut {NoStop}%
\bibitem [{\citenamefont {Yan}\ and\ \citenamefont
  {Felser}(2017)}]{Felser_ARCMP}%
  \BibitemOpen
  \bibfield  {author} {\bibinfo {author} {\bibfnamefont {B.}~\bibnamefont
  {Yan}}\ and\ \bibinfo {author} {\bibfnamefont {C.}~\bibnamefont {Felser}},\
  }\href@noop {} {\bibfield  {journal} {\bibinfo  {journal} {Annual Review of
  Condensed Matter Physics}\ }\textbf {\bibinfo {volume} {8}} (\bibinfo {year}
  {2017})}\BibitemShut {NoStop}%
\bibitem [{\citenamefont {{Burkov}}(2018)}]{Burkov_ARCMP}%
  \BibitemOpen
  \bibfield  {author} {\bibinfo {author} {\bibfnamefont {A.~A.}\ \bibnamefont
  {{Burkov}}},\ }\href@noop {} {\bibfield  {journal} {\bibinfo  {journal}
  {Annual Review of Condensed Matter Physics}\ }\textbf {\bibinfo {volume}
  {9}},\ \bibinfo {pages} {359} (\bibinfo {year} {2018})}\BibitemShut {NoStop}%
\bibitem [{\citenamefont {Wan}\ \emph {et~al.}(2011)\citenamefont {Wan},
  \citenamefont {Turner}, \citenamefont {Vishwanath},\ and\ \citenamefont
  {Savrasov}}]{Wan11}%
  \BibitemOpen
  \bibfield  {author} {\bibinfo {author} {\bibfnamefont {X.}~\bibnamefont
  {Wan}}, \bibinfo {author} {\bibfnamefont {A.~M.}\ \bibnamefont {Turner}},
  \bibinfo {author} {\bibfnamefont {A.}~\bibnamefont {Vishwanath}}, \ and\
  \bibinfo {author} {\bibfnamefont {S.~Y.}\ \bibnamefont {Savrasov}},\ }\href
  {\doibase 10.1103/PhysRevB.83.205101} {\bibfield  {journal} {\bibinfo
  {journal} {Phys. Rev. B}\ }\textbf {\bibinfo {volume} {83}},\ \bibinfo
  {pages} {205101} (\bibinfo {year} {2011})}\BibitemShut {NoStop}%
\bibitem [{\citenamefont {Burkov}\ and\ \citenamefont
  {Balents}(2011)}]{Burkov11-1}%
  \BibitemOpen
  \bibfield  {author} {\bibinfo {author} {\bibfnamefont {A.~A.}\ \bibnamefont
  {Burkov}}\ and\ \bibinfo {author} {\bibfnamefont {L.}~\bibnamefont
  {Balents}},\ }\href {\doibase 10.1103/PhysRevLett.107.127205} {\bibfield
  {journal} {\bibinfo  {journal} {Phys. Rev. Lett.}\ }\textbf {\bibinfo
  {volume} {107}},\ \bibinfo {pages} {127205} (\bibinfo {year}
  {2011})}\BibitemShut {NoStop}%
\bibitem [{\citenamefont {Burkov}\ \emph {et~al.}(2011)\citenamefont {Burkov},
  \citenamefont {Hook},\ and\ \citenamefont {Balents}}]{Burkov11-2}%
  \BibitemOpen
  \bibfield  {author} {\bibinfo {author} {\bibfnamefont {A.~A.}\ \bibnamefont
  {Burkov}}, \bibinfo {author} {\bibfnamefont {M.~D.}\ \bibnamefont {Hook}}, \
  and\ \bibinfo {author} {\bibfnamefont {L.}~\bibnamefont {Balents}},\ }\href
  {\doibase 10.1103/PhysRevB.84.235126} {\bibfield  {journal} {\bibinfo
  {journal} {Phys. Rev. B}\ }\textbf {\bibinfo {volume} {84}},\ \bibinfo
  {pages} {235126} (\bibinfo {year} {2011})}\BibitemShut {NoStop}%
\bibitem [{\citenamefont {Xu}\ \emph {et~al.}(2011)\citenamefont {Xu},
  \citenamefont {Weng}, \citenamefont {Wang}, \citenamefont {Dai},\ and\
  \citenamefont {Fang}}]{Xu11}%
  \BibitemOpen
  \bibfield  {author} {\bibinfo {author} {\bibfnamefont {G.}~\bibnamefont
  {Xu}}, \bibinfo {author} {\bibfnamefont {H.}~\bibnamefont {Weng}}, \bibinfo
  {author} {\bibfnamefont {Z.}~\bibnamefont {Wang}}, \bibinfo {author}
  {\bibfnamefont {X.}~\bibnamefont {Dai}}, \ and\ \bibinfo {author}
  {\bibfnamefont {Z.}~\bibnamefont {Fang}},\ }\href {\doibase
  10.1103/PhysRevLett.107.186806} {\bibfield  {journal} {\bibinfo  {journal}
  {Phys. Rev. Lett.}\ }\textbf {\bibinfo {volume} {107}},\ \bibinfo {pages}
  {186806} (\bibinfo {year} {2011})}\BibitemShut {NoStop}%
\bibitem [{\citenamefont {Young}\ \emph {et~al.}(2012)\citenamefont {Young},
  \citenamefont {Zaheer}, \citenamefont {Teo}, \citenamefont {Kane},
  \citenamefont {Mele},\ and\ \citenamefont {Rappe}}]{Kane12}%
  \BibitemOpen
  \bibfield  {author} {\bibinfo {author} {\bibfnamefont {S.~M.}\ \bibnamefont
  {Young}}, \bibinfo {author} {\bibfnamefont {S.}~\bibnamefont {Zaheer}},
  \bibinfo {author} {\bibfnamefont {J.~C.~Y.}\ \bibnamefont {Teo}}, \bibinfo
  {author} {\bibfnamefont {C.~L.}\ \bibnamefont {Kane}}, \bibinfo {author}
  {\bibfnamefont {E.~J.}\ \bibnamefont {Mele}}, \ and\ \bibinfo {author}
  {\bibfnamefont {A.~M.}\ \bibnamefont {Rappe}},\ }\href {\doibase
  10.1103/PhysRevLett.108.140405} {\bibfield  {journal} {\bibinfo  {journal}
  {Phys. Rev. Lett.}\ }\textbf {\bibinfo {volume} {108}},\ \bibinfo {pages}
  {140405} (\bibinfo {year} {2012})}\BibitemShut {NoStop}%
\bibitem [{\citenamefont {Wang}\ \emph {et~al.}(2012)\citenamefont {Wang},
  \citenamefont {Sun}, \citenamefont {Chen}, \citenamefont {Franchini},
  \citenamefont {Xu}, \citenamefont {Weng}, \citenamefont {Dai},\ and\
  \citenamefont {Fang}}]{Fang12}%
  \BibitemOpen
  \bibfield  {author} {\bibinfo {author} {\bibfnamefont {Z.}~\bibnamefont
  {Wang}}, \bibinfo {author} {\bibfnamefont {Y.}~\bibnamefont {Sun}}, \bibinfo
  {author} {\bibfnamefont {X.-Q.}\ \bibnamefont {Chen}}, \bibinfo {author}
  {\bibfnamefont {C.}~\bibnamefont {Franchini}}, \bibinfo {author}
  {\bibfnamefont {G.}~\bibnamefont {Xu}}, \bibinfo {author} {\bibfnamefont
  {H.}~\bibnamefont {Weng}}, \bibinfo {author} {\bibfnamefont {X.}~\bibnamefont
  {Dai}}, \ and\ \bibinfo {author} {\bibfnamefont {Z.}~\bibnamefont {Fang}},\
  }\href {\doibase 10.1103/PhysRevB.85.195320} {\bibfield  {journal} {\bibinfo
  {journal} {Phys. Rev. B}\ }\textbf {\bibinfo {volume} {85}},\ \bibinfo
  {pages} {195320} (\bibinfo {year} {2012})}\BibitemShut {NoStop}%
\bibitem [{\citenamefont {Wang}\ \emph {et~al.}(2013)\citenamefont {Wang},
  \citenamefont {Weng}, \citenamefont {Wu}, \citenamefont {Dai},\ and\
  \citenamefont {Fang}}]{Fang13}%
  \BibitemOpen
  \bibfield  {author} {\bibinfo {author} {\bibfnamefont {Z.}~\bibnamefont
  {Wang}}, \bibinfo {author} {\bibfnamefont {H.}~\bibnamefont {Weng}}, \bibinfo
  {author} {\bibfnamefont {Q.}~\bibnamefont {Wu}}, \bibinfo {author}
  {\bibfnamefont {X.}~\bibnamefont {Dai}}, \ and\ \bibinfo {author}
  {\bibfnamefont {Z.}~\bibnamefont {Fang}},\ }\href {\doibase
  10.1103/PhysRevB.88.125427} {\bibfield  {journal} {\bibinfo  {journal} {Phys.
  Rev. B}\ }\textbf {\bibinfo {volume} {88}},\ \bibinfo {pages} {125427}
  (\bibinfo {year} {2013})}\BibitemShut {NoStop}%
\bibitem [{\citenamefont {Liu}\ \emph {et~al.}(2014)\citenamefont {Liu},
  \citenamefont {Zhou}, \citenamefont {Zhang}, \citenamefont {Wang},
  \citenamefont {Weng}, \citenamefont {Prabhakaran}, \citenamefont {Mo},
  \citenamefont {Shen}, \citenamefont {Fang}, \citenamefont {Dai},
  \citenamefont {Hussain},\ and\ \citenamefont {Chen}}]{Chen14}%
  \BibitemOpen
  \bibfield  {author} {\bibinfo {author} {\bibfnamefont {Z.~K.}\ \bibnamefont
  {Liu}}, \bibinfo {author} {\bibfnamefont {B.}~\bibnamefont {Zhou}}, \bibinfo
  {author} {\bibfnamefont {Y.}~\bibnamefont {Zhang}}, \bibinfo {author}
  {\bibfnamefont {Z.~J.}\ \bibnamefont {Wang}}, \bibinfo {author}
  {\bibfnamefont {H.~M.}\ \bibnamefont {Weng}}, \bibinfo {author}
  {\bibfnamefont {D.}~\bibnamefont {Prabhakaran}}, \bibinfo {author}
  {\bibfnamefont {S.-K.}\ \bibnamefont {Mo}}, \bibinfo {author} {\bibfnamefont
  {Z.~X.}\ \bibnamefont {Shen}}, \bibinfo {author} {\bibfnamefont
  {Z.}~\bibnamefont {Fang}}, \bibinfo {author} {\bibfnamefont {X.}~\bibnamefont
  {Dai}}, \bibinfo {author} {\bibfnamefont {Z.}~\bibnamefont {Hussain}}, \ and\
  \bibinfo {author} {\bibfnamefont {Y.~L.}\ \bibnamefont {Chen}},\ }\href
  {\doibase 10.1126/science.1245085} {\bibfield  {journal} {\bibinfo  {journal}
  {Science}\ }\textbf {\bibinfo {volume} {343}},\ \bibinfo {pages} {864}
  (\bibinfo {year} {2014})}\BibitemShut {NoStop}%
\bibitem [{\citenamefont {Neupane}\ \emph {et~al.}(2014)\citenamefont
  {Neupane}, \citenamefont {Xu}, \citenamefont {Sankar}, \citenamefont
  {Alidoust}, \citenamefont {Bian}, \citenamefont {Liu}, \citenamefont
  {Belopolski}, \citenamefont {Chang}, \citenamefont {Jeng}, \citenamefont
  {Lin}, \citenamefont {Bansil}, \citenamefont {Chou},\ and\ \citenamefont
  {Hasan}}]{Neupane14}%
  \BibitemOpen
  \bibfield  {author} {\bibinfo {author} {\bibfnamefont {M.}~\bibnamefont
  {Neupane}}, \bibinfo {author} {\bibfnamefont {S.-Y.}\ \bibnamefont {Xu}},
  \bibinfo {author} {\bibfnamefont {R.}~\bibnamefont {Sankar}}, \bibinfo
  {author} {\bibfnamefont {N.}~\bibnamefont {Alidoust}}, \bibinfo {author}
  {\bibfnamefont {G.}~\bibnamefont {Bian}}, \bibinfo {author} {\bibfnamefont
  {C.}~\bibnamefont {Liu}}, \bibinfo {author} {\bibfnamefont {I.}~\bibnamefont
  {Belopolski}}, \bibinfo {author} {\bibfnamefont {T.-R.}\ \bibnamefont
  {Chang}}, \bibinfo {author} {\bibfnamefont {H.-T.}\ \bibnamefont {Jeng}},
  \bibinfo {author} {\bibfnamefont {H.}~\bibnamefont {Lin}}, \bibinfo {author}
  {\bibfnamefont {A.}~\bibnamefont {Bansil}}, \bibinfo {author} {\bibfnamefont
  {F.}~\bibnamefont {Chou}}, \ and\ \bibinfo {author} {\bibfnamefont {M.~Z.}\
  \bibnamefont {Hasan}},\ }\href {http://dx.doi.org/10.1038/ncomms4786}
  {\bibfield  {journal} {\bibinfo  {journal} {Nat. Commun.}\ }\textbf {\bibinfo
  {volume} {5}} (\bibinfo {year} {2014})}\BibitemShut {NoStop}%
\bibitem [{\citenamefont {Xu}\ \emph {et~al.}(2015)\citenamefont {Xu},
  \citenamefont {Belopolski}, \citenamefont {Alidoust}, \citenamefont
  {Neupane}, \citenamefont {Bian}, \citenamefont {Zhang}, \citenamefont
  {Sankar}, \citenamefont {Chang}, \citenamefont {Yuan}, \citenamefont {Lee},
  \citenamefont {Huang}, \citenamefont {Zheng}, \citenamefont {Ma},
  \citenamefont {Sanchez}, \citenamefont {Wang}, \citenamefont {Bansil},
  \citenamefont {Chou}, \citenamefont {Shibayev}, \citenamefont {Lin},
  \citenamefont {Jia},\ and\ \citenamefont {Hasan}}]{HasanTaAs}%
  \BibitemOpen
  \bibfield  {author} {\bibinfo {author} {\bibfnamefont {S.-Y.}\ \bibnamefont
  {Xu}}, \bibinfo {author} {\bibfnamefont {I.}~\bibnamefont {Belopolski}},
  \bibinfo {author} {\bibfnamefont {N.}~\bibnamefont {Alidoust}}, \bibinfo
  {author} {\bibfnamefont {M.}~\bibnamefont {Neupane}}, \bibinfo {author}
  {\bibfnamefont {G.}~\bibnamefont {Bian}}, \bibinfo {author} {\bibfnamefont
  {C.}~\bibnamefont {Zhang}}, \bibinfo {author} {\bibfnamefont
  {R.}~\bibnamefont {Sankar}}, \bibinfo {author} {\bibfnamefont
  {G.}~\bibnamefont {Chang}}, \bibinfo {author} {\bibfnamefont
  {Z.}~\bibnamefont {Yuan}}, \bibinfo {author} {\bibfnamefont {C.-C.}\
  \bibnamefont {Lee}}, \bibinfo {author} {\bibfnamefont {S.-M.}\ \bibnamefont
  {Huang}}, \bibinfo {author} {\bibfnamefont {H.}~\bibnamefont {Zheng}},
  \bibinfo {author} {\bibfnamefont {J.}~\bibnamefont {Ma}}, \bibinfo {author}
  {\bibfnamefont {D.~S.}\ \bibnamefont {Sanchez}}, \bibinfo {author}
  {\bibfnamefont {B.}~\bibnamefont {Wang}}, \bibinfo {author} {\bibfnamefont
  {A.}~\bibnamefont {Bansil}}, \bibinfo {author} {\bibfnamefont
  {F.}~\bibnamefont {Chou}}, \bibinfo {author} {\bibfnamefont {P.~P.}\
  \bibnamefont {Shibayev}}, \bibinfo {author} {\bibfnamefont {H.}~\bibnamefont
  {Lin}}, \bibinfo {author} {\bibfnamefont {S.}~\bibnamefont {Jia}}, \ and\
  \bibinfo {author} {\bibfnamefont {M.~Z.}\ \bibnamefont {Hasan}},\ }\href
  {\doibase 10.1126/science.aaa9297} {\bibfield  {journal} {\bibinfo  {journal}
  {Science}\ }\textbf {\bibinfo {volume} {349}},\ \bibinfo {pages} {613}
  (\bibinfo {year} {2015})}\BibitemShut {NoStop}%
\bibitem [{\citenamefont {Lv}\ \emph {et~al.}(2015{\natexlab{a}})\citenamefont
  {Lv}, \citenamefont {Xu}, \citenamefont {Weng}, \citenamefont {Ma},
  \citenamefont {Richard}, \citenamefont {Huang}, \citenamefont {Zhao},
  \citenamefont {Chen}, \citenamefont {Matt}, \citenamefont {Bisti},
  \citenamefont {Strocov}, \citenamefont {Mesot}, \citenamefont {Fang},
  \citenamefont {Dai}, \citenamefont {Qian}, \citenamefont {Shi},\ and\
  \citenamefont {Ding}}]{DingTaAs2}%
  \BibitemOpen
  \bibfield  {author} {\bibinfo {author} {\bibfnamefont {B.~Q.}\ \bibnamefont
  {Lv}}, \bibinfo {author} {\bibfnamefont {N.}~\bibnamefont {Xu}}, \bibinfo
  {author} {\bibfnamefont {H.~M.}\ \bibnamefont {Weng}}, \bibinfo {author}
  {\bibfnamefont {J.~Z.}\ \bibnamefont {Ma}}, \bibinfo {author} {\bibfnamefont
  {P.}~\bibnamefont {Richard}}, \bibinfo {author} {\bibfnamefont {X.~C.}\
  \bibnamefont {Huang}}, \bibinfo {author} {\bibfnamefont {L.~X.}\ \bibnamefont
  {Zhao}}, \bibinfo {author} {\bibfnamefont {G.~F.}\ \bibnamefont {Chen}},
  \bibinfo {author} {\bibfnamefont {C.~E.}\ \bibnamefont {Matt}}, \bibinfo
  {author} {\bibfnamefont {F.}~\bibnamefont {Bisti}}, \bibinfo {author}
  {\bibfnamefont {V.~N.}\ \bibnamefont {Strocov}}, \bibinfo {author}
  {\bibfnamefont {J.}~\bibnamefont {Mesot}}, \bibinfo {author} {\bibfnamefont
  {Z.}~\bibnamefont {Fang}}, \bibinfo {author} {\bibfnamefont {X.}~\bibnamefont
  {Dai}}, \bibinfo {author} {\bibfnamefont {T.}~\bibnamefont {Qian}}, \bibinfo
  {author} {\bibfnamefont {M.}~\bibnamefont {Shi}}, \ and\ \bibinfo {author}
  {\bibfnamefont {H.}~\bibnamefont {Ding}},\ }\href
  {http://dx.doi.org/10.1038/nphys3426} {\bibfield  {journal} {\bibinfo
  {journal} {Nat Phys}\ }\textbf {\bibinfo {volume} {11}},\ \bibinfo {pages}
  {724} (\bibinfo {year} {2015}{\natexlab{a}})}\BibitemShut {NoStop}%
\bibitem [{\citenamefont {Lv}\ \emph {et~al.}(2015{\natexlab{b}})\citenamefont
  {Lv}, \citenamefont {Weng}, \citenamefont {Fu}, \citenamefont {Wang},
  \citenamefont {Miao}, \citenamefont {Ma}, \citenamefont {Richard},
  \citenamefont {Huang}, \citenamefont {Zhao}, \citenamefont {Chen},
  \citenamefont {Fang}, \citenamefont {Dai}, \citenamefont {Qian},\ and\
  \citenamefont {Ding}}]{DingTaAs}%
  \BibitemOpen
  \bibfield  {author} {\bibinfo {author} {\bibfnamefont {B.~Q.}\ \bibnamefont
  {Lv}}, \bibinfo {author} {\bibfnamefont {H.~M.}\ \bibnamefont {Weng}},
  \bibinfo {author} {\bibfnamefont {B.~B.}\ \bibnamefont {Fu}}, \bibinfo
  {author} {\bibfnamefont {X.~P.}\ \bibnamefont {Wang}}, \bibinfo {author}
  {\bibfnamefont {H.}~\bibnamefont {Miao}}, \bibinfo {author} {\bibfnamefont
  {J.}~\bibnamefont {Ma}}, \bibinfo {author} {\bibfnamefont {P.}~\bibnamefont
  {Richard}}, \bibinfo {author} {\bibfnamefont {X.~C.}\ \bibnamefont {Huang}},
  \bibinfo {author} {\bibfnamefont {L.~X.}\ \bibnamefont {Zhao}}, \bibinfo
  {author} {\bibfnamefont {G.~F.}\ \bibnamefont {Chen}}, \bibinfo {author}
  {\bibfnamefont {Z.}~\bibnamefont {Fang}}, \bibinfo {author} {\bibfnamefont
  {X.}~\bibnamefont {Dai}}, \bibinfo {author} {\bibfnamefont {T.}~\bibnamefont
  {Qian}}, \ and\ \bibinfo {author} {\bibfnamefont {H.}~\bibnamefont {Ding}},\
  }\href {\doibase 10.1103/PhysRevX.5.031013} {\bibfield  {journal} {\bibinfo
  {journal} {Phys. Rev. X}\ }\textbf {\bibinfo {volume} {5}},\ \bibinfo {pages}
  {031013} (\bibinfo {year} {2015}{\natexlab{b}})}\BibitemShut {NoStop}%
\bibitem [{\citenamefont {Lu}\ \emph {et~al.}(2015)\citenamefont {Lu},
  \citenamefont {Wang}, \citenamefont {Ye}, \citenamefont {Ran}, \citenamefont
  {Fu}, \citenamefont {Joannopoulos},\ and\ \citenamefont {Solja{\v
  c}i{\'c}}}]{Lu15}%
  \BibitemOpen
  \bibfield  {author} {\bibinfo {author} {\bibfnamefont {L.}~\bibnamefont
  {Lu}}, \bibinfo {author} {\bibfnamefont {Z.}~\bibnamefont {Wang}}, \bibinfo
  {author} {\bibfnamefont {D.}~\bibnamefont {Ye}}, \bibinfo {author}
  {\bibfnamefont {L.}~\bibnamefont {Ran}}, \bibinfo {author} {\bibfnamefont
  {L.}~\bibnamefont {Fu}}, \bibinfo {author} {\bibfnamefont {J.~D.}\
  \bibnamefont {Joannopoulos}}, \ and\ \bibinfo {author} {\bibfnamefont
  {M.}~\bibnamefont {Solja{\v c}i{\'c}}},\ }\href {\doibase
  10.1126/science.aaa9273} {\bibfield  {journal} {\bibinfo  {journal}
  {Science}\ }\textbf {\bibinfo {volume} {349}},\ \bibinfo {pages} {622}
  (\bibinfo {year} {2015})}\BibitemShut {NoStop}%
\bibitem [{\citenamefont {{Liu}}\ \emph {et~al.}(2017)\citenamefont {{Liu}},
  \citenamefont {{Sun}}, \citenamefont {{M{\"u}echler}}, \citenamefont {{Sun}},
  \citenamefont {{Jiao}}, \citenamefont {{Kroder}}, \citenamefont
  {{S{\"u}{\ss}}}, \citenamefont {{Borrmann}}, \citenamefont {{Wang}},
  \citenamefont {{Schnelle}}, \citenamefont {{Wirth}}, \citenamefont
  {{Goennenwein}},\ and\ \citenamefont {{Felser}}}]{Felser17}%
  \BibitemOpen
  \bibfield  {author} {\bibinfo {author} {\bibfnamefont {E.}~\bibnamefont
  {{Liu}}}, \bibinfo {author} {\bibfnamefont {Y.}~\bibnamefont {{Sun}}},
  \bibinfo {author} {\bibfnamefont {L.}~\bibnamefont {{M{\"u}echler}}},
  \bibinfo {author} {\bibfnamefont {A.}~\bibnamefont {{Sun}}}, \bibinfo
  {author} {\bibfnamefont {L.}~\bibnamefont {{Jiao}}}, \bibinfo {author}
  {\bibfnamefont {J.}~\bibnamefont {{Kroder}}}, \bibinfo {author}
  {\bibfnamefont {V.}~\bibnamefont {{S{\"u}{\ss}}}}, \bibinfo {author}
  {\bibfnamefont {H.}~\bibnamefont {{Borrmann}}}, \bibinfo {author}
  {\bibfnamefont {W.}~\bibnamefont {{Wang}}}, \bibinfo {author} {\bibfnamefont
  {W.}~\bibnamefont {{Schnelle}}}, \bibinfo {author} {\bibfnamefont
  {S.}~\bibnamefont {{Wirth}}}, \bibinfo {author} {\bibfnamefont {S.~T.~B.}\
  \bibnamefont {{Goennenwein}}}, \ and\ \bibinfo {author} {\bibfnamefont
  {C.}~\bibnamefont {{Felser}}},\ }\href@noop {} {\bibfield  {journal}
  {\bibinfo  {journal} {ArXiv e-prints}\ } (\bibinfo {year} {2017})},\ \Eprint
  {http://arxiv.org/abs/1712.06722} {arXiv:1712.06722 [cond-mat.mtrl-sci]}
  \BibitemShut {NoStop}%
\bibitem [{\citenamefont {Volovik}(2003)}]{Volovik03}%
  \BibitemOpen
  \bibfield  {author} {\bibinfo {author} {\bibfnamefont {G.}~\bibnamefont
  {Volovik}},\ }\href@noop {} {\emph {\bibinfo {title} {The Universe in a
  Helium Droplet}}}\ (\bibinfo  {publisher} {Oxford: Clarendon},\ \bibinfo
  {year} {2003})\BibitemShut {NoStop}%
\bibitem [{\citenamefont {Haldane}(2004)}]{Haldane04}%
  \BibitemOpen
  \bibfield  {author} {\bibinfo {author} {\bibfnamefont {F.~D.~M.}\
  \bibnamefont {Haldane}},\ }\href {\doibase 10.1103/PhysRevLett.93.206602}
  {\bibfield  {journal} {\bibinfo  {journal} {Phys. Rev. Lett.}\ }\textbf
  {\bibinfo {volume} {93}},\ \bibinfo {pages} {206602} (\bibinfo {year}
  {2004})}\BibitemShut {NoStop}%
\bibitem [{\citenamefont {Volovik}(2007)}]{Volovik07}%
  \BibitemOpen
  \bibfield  {author} {\bibinfo {author} {\bibfnamefont {G.~E.}\ \bibnamefont
  {Volovik}},\ }in\ \href {\doibase 10.1007/3-540-70859-6_3} {\emph {\bibinfo
  {booktitle} {Quantum Analogues: From Phase Transitions to Black Holes and
  Cosmology}}},\ \bibinfo {series} {Lecture Notes in Physics}, Vol.\ \bibinfo
  {volume} {718},\ \bibinfo {editor} {edited by\ \bibinfo {editor}
  {\bibfnamefont {W.}~\bibnamefont {Unruh}}\ and\ \bibinfo {editor}
  {\bibfnamefont {R.}~\bibnamefont {Schützhold}}}\ (\bibinfo  {publisher}
  {Springer Berlin Heidelberg},\ \bibinfo {year} {2007})\BibitemShut {NoStop}%
\bibitem [{\citenamefont {Murakami}(2007)}]{Murakami07}%
  \BibitemOpen
  \bibfield  {author} {\bibinfo {author} {\bibfnamefont {S.}~\bibnamefont
  {Murakami}},\ }\href {http://stacks.iop.org/1367-2630/9/i=9/a=356} {\bibfield
   {journal} {\bibinfo  {journal} {New Journal of Physics}\ }\textbf {\bibinfo
  {volume} {9}},\ \bibinfo {pages} {356} (\bibinfo {year} {2007})}\BibitemShut
  {NoStop}%
\bibitem [{\citenamefont {Zyuzin}\ and\ \citenamefont
  {Burkov}(2012)}]{Zyuzin12-1}%
  \BibitemOpen
  \bibfield  {author} {\bibinfo {author} {\bibfnamefont {A.~A.}\ \bibnamefont
  {Zyuzin}}\ and\ \bibinfo {author} {\bibfnamefont {A.~A.}\ \bibnamefont
  {Burkov}},\ }\href {\doibase 10.1103/PhysRevB.86.115133} {\bibfield
  {journal} {\bibinfo  {journal} {Phys. Rev. B}\ }\textbf {\bibinfo {volume}
  {86}},\ \bibinfo {pages} {115133} (\bibinfo {year} {2012})}\BibitemShut
  {NoStop}%
\bibitem [{\citenamefont {Nielsen}\ and\ \citenamefont
  {Ninomiya}(1983)}]{Nielsen83}%
  \BibitemOpen
  \bibfield  {author} {\bibinfo {author} {\bibfnamefont {H.}~\bibnamefont
  {Nielsen}}\ and\ \bibinfo {author} {\bibfnamefont {M.}~\bibnamefont
  {Ninomiya}},\ }\href {\doibase
  http://dx.doi.org/10.1016/0370-2693(83)91529-0} {\bibfield  {journal}
  {\bibinfo  {journal} {Physics Letters B}\ }\textbf {\bibinfo {volume}
  {130}},\ \bibinfo {pages} {389 } (\bibinfo {year} {1983})}\BibitemShut
  {NoStop}%
\bibitem [{\citenamefont {Son}\ and\ \citenamefont {Spivak}(2013)}]{Spivak12}%
  \BibitemOpen
  \bibfield  {author} {\bibinfo {author} {\bibfnamefont {D.~T.}\ \bibnamefont
  {Son}}\ and\ \bibinfo {author} {\bibfnamefont {B.~Z.}\ \bibnamefont
  {Spivak}},\ }\href {\doibase 10.1103/PhysRevB.88.104412} {\bibfield
  {journal} {\bibinfo  {journal} {Phys. Rev. B}\ }\textbf {\bibinfo {volume}
  {88}},\ \bibinfo {pages} {104412} (\bibinfo {year} {2013})}\BibitemShut
  {NoStop}%
\bibitem [{\citenamefont {Burkov}(2015)}]{Burkov_lmr_prb}%
  \BibitemOpen
  \bibfield  {author} {\bibinfo {author} {\bibfnamefont {A.~A.}\ \bibnamefont
  {Burkov}},\ }\href {\doibase 10.1103/PhysRevB.91.245157} {\bibfield
  {journal} {\bibinfo  {journal} {Phys. Rev. B}\ }\textbf {\bibinfo {volume}
  {91}},\ \bibinfo {pages} {245157} (\bibinfo {year} {2015})}\BibitemShut
  {NoStop}%
\bibitem [{\citenamefont {Burkov}(2017)}]{Burkov_gphe}%
  \BibitemOpen
  \bibfield  {author} {\bibinfo {author} {\bibfnamefont {A.~A.}\ \bibnamefont
  {Burkov}},\ }\href {\doibase 10.1103/PhysRevB.96.041110} {\bibfield
  {journal} {\bibinfo  {journal} {Phys. Rev. B}\ }\textbf {\bibinfo {volume}
  {96}},\ \bibinfo {pages} {041110} (\bibinfo {year} {2017})}\BibitemShut
  {NoStop}%
\bibitem [{\citenamefont {Nandy}\ \emph {et~al.}(2017)\citenamefont {Nandy},
  \citenamefont {Sharma}, \citenamefont {Taraphder},\ and\ \citenamefont
  {Tewari}}]{Tewari_gphe}%
  \BibitemOpen
  \bibfield  {author} {\bibinfo {author} {\bibfnamefont {S.}~\bibnamefont
  {Nandy}}, \bibinfo {author} {\bibfnamefont {G.}~\bibnamefont {Sharma}},
  \bibinfo {author} {\bibfnamefont {A.}~\bibnamefont {Taraphder}}, \ and\
  \bibinfo {author} {\bibfnamefont {S.}~\bibnamefont {Tewari}},\ }\href
  {\doibase 10.1103/PhysRevLett.119.176804} {\bibfield  {journal} {\bibinfo
  {journal} {Phys. Rev. Lett.}\ }\textbf {\bibinfo {volume} {119}},\ \bibinfo
  {pages} {176804} (\bibinfo {year} {2017})}\BibitemShut {NoStop}%
\bibitem [{\citenamefont {Burkov}(2014)}]{Burkov_AHE}%
  \BibitemOpen
  \bibfield  {author} {\bibinfo {author} {\bibfnamefont {A.~A.}\ \bibnamefont
  {Burkov}},\ }\href {\doibase 10.1103/PhysRevLett.113.187202} {\bibfield
  {journal} {\bibinfo  {journal} {Phys. Rev. Lett.}\ }\textbf {\bibinfo
  {volume} {113}},\ \bibinfo {pages} {187202} (\bibinfo {year}
  {2014})}\BibitemShut {NoStop}%
\bibitem [{\citenamefont {Xiong}\ \emph {et~al.}(2015)\citenamefont {Xiong},
  \citenamefont {Kushwaha}, \citenamefont {Liang}, \citenamefont {Krizan},
  \citenamefont {Hirschberger}, \citenamefont {Wang}, \citenamefont {Cava},\
  and\ \citenamefont {Ong}}]{Ong_anomaly}%
  \BibitemOpen
  \bibfield  {author} {\bibinfo {author} {\bibfnamefont {J.}~\bibnamefont
  {Xiong}}, \bibinfo {author} {\bibfnamefont {S.~K.}\ \bibnamefont {Kushwaha}},
  \bibinfo {author} {\bibfnamefont {T.}~\bibnamefont {Liang}}, \bibinfo
  {author} {\bibfnamefont {J.~W.}\ \bibnamefont {Krizan}}, \bibinfo {author}
  {\bibfnamefont {M.}~\bibnamefont {Hirschberger}}, \bibinfo {author}
  {\bibfnamefont {W.}~\bibnamefont {Wang}}, \bibinfo {author} {\bibfnamefont
  {R.~J.}\ \bibnamefont {Cava}}, \ and\ \bibinfo {author} {\bibfnamefont
  {N.~P.}\ \bibnamefont {Ong}},\ }\href {\doibase 10.1126/science.aac6089}
  {\bibfield  {journal} {\bibinfo  {journal} {Science}\ }\textbf {\bibinfo
  {volume} {350}},\ \bibinfo {pages} {413} (\bibinfo {year}
  {2015})}\BibitemShut {NoStop}%
\bibitem [{\citenamefont {Li}\ \emph {et~al.}(2016)\citenamefont {Li},
  \citenamefont {Kharzeev}, \citenamefont {Zhang}, \citenamefont {Huang},
  \citenamefont {Pletikosic}, \citenamefont {Fedorov}, \citenamefont {Zhong},
  \citenamefont {Schneeloch}, \citenamefont {Gu},\ and\ \citenamefont
  {Valla}}]{Li_anomaly}%
  \BibitemOpen
  \bibfield  {author} {\bibinfo {author} {\bibfnamefont {Q.}~\bibnamefont
  {Li}}, \bibinfo {author} {\bibfnamefont {D.~E.}\ \bibnamefont {Kharzeev}},
  \bibinfo {author} {\bibfnamefont {C.}~\bibnamefont {Zhang}}, \bibinfo
  {author} {\bibfnamefont {Y.}~\bibnamefont {Huang}}, \bibinfo {author}
  {\bibfnamefont {I.}~\bibnamefont {Pletikosic}}, \bibinfo {author}
  {\bibfnamefont {A.~V.}\ \bibnamefont {Fedorov}}, \bibinfo {author}
  {\bibfnamefont {R.~D.}\ \bibnamefont {Zhong}}, \bibinfo {author}
  {\bibfnamefont {J.~A.}\ \bibnamefont {Schneeloch}}, \bibinfo {author}
  {\bibfnamefont {G.~D.}\ \bibnamefont {Gu}}, \ and\ \bibinfo {author}
  {\bibfnamefont {T.}~\bibnamefont {Valla}},\ }\href
  {http://dx.doi.org/10.1038/nphys3648} {\bibfield  {journal} {\bibinfo
  {journal} {Nat Phys}\ }\textbf {\bibinfo {volume} {12}},\ \bibinfo {pages}
  {550} (\bibinfo {year} {2016})}\BibitemShut {NoStop}%
\bibitem [{\citenamefont {{Li}}\ \emph {et~al.}(2017)\citenamefont {{Li}},
  \citenamefont {{Wang}}, \citenamefont {{He}}, \citenamefont {{Wang}},\ and\
  \citenamefont {{Shen}}}]{Shen_gphe}%
  \BibitemOpen
  \bibfield  {author} {\bibinfo {author} {\bibfnamefont {H.}~\bibnamefont
  {{Li}}}, \bibinfo {author} {\bibfnamefont {H.}~\bibnamefont {{Wang}}},
  \bibinfo {author} {\bibfnamefont {H.}~\bibnamefont {{He}}}, \bibinfo {author}
  {\bibfnamefont {J.}~\bibnamefont {{Wang}}}, \ and\ \bibinfo {author}
  {\bibfnamefont {S.-Q.}\ \bibnamefont {{Shen}}},\ }\href@noop {} {\bibfield
  {journal} {\bibinfo  {journal} {ArXiv e-prints}\ } (\bibinfo {year}
  {2017})},\ \Eprint {http://arxiv.org/abs/1711.03671} {arXiv:1711.03671
  [cond-mat.mes-hall]} \BibitemShut {NoStop}%
\bibitem [{\citenamefont {{Kumar}}\ \emph {et~al.}(2017)\citenamefont
  {{Kumar}}, \citenamefont {{Felser}},\ and\ \citenamefont
  {{Shekhar}}}]{Felser_gphe}%
  \BibitemOpen
  \bibfield  {author} {\bibinfo {author} {\bibfnamefont {N.}~\bibnamefont
  {{Kumar}}}, \bibinfo {author} {\bibfnamefont {C.}~\bibnamefont {{Felser}}}, \
  and\ \bibinfo {author} {\bibfnamefont {C.}~\bibnamefont {{Shekhar}}},\
  }\href@noop {} {\bibfield  {journal} {\bibinfo  {journal} {ArXiv e-prints}\ }
  (\bibinfo {year} {2017})},\ \Eprint {http://arxiv.org/abs/1711.04133}
  {arXiv:1711.04133 [cond-mat.mes-hall]} \BibitemShut {NoStop}%
\bibitem [{\citenamefont {{Wang}}\ \emph {et~al.}(2018)\citenamefont {{Wang}},
  \citenamefont {{Gong}}, \citenamefont {{Liang}}, \citenamefont {{Ge}},
  \citenamefont {{Wang}}, \citenamefont {{Zhu}},\ and\ \citenamefont
  {{Zhang}}}]{Zhang_gphe}%
  \BibitemOpen
  \bibfield  {author} {\bibinfo {author} {\bibfnamefont {Y.~J.}\ \bibnamefont
  {{Wang}}}, \bibinfo {author} {\bibfnamefont {J.~X.}\ \bibnamefont {{Gong}}},
  \bibinfo {author} {\bibfnamefont {D.~D.}\ \bibnamefont {{Liang}}}, \bibinfo
  {author} {\bibfnamefont {M.}~\bibnamefont {{Ge}}}, \bibinfo {author}
  {\bibfnamefont {J.~R.}\ \bibnamefont {{Wang}}}, \bibinfo {author}
  {\bibfnamefont {W.~K.}\ \bibnamefont {{Zhu}}}, \ and\ \bibinfo {author}
  {\bibfnamefont {C.~J.}\ \bibnamefont {{Zhang}}},\ }\href@noop {} {\bibfield
  {journal} {\bibinfo  {journal} {ArXiv e-prints}\ } (\bibinfo {year}
  {2018})},\ \Eprint {http://arxiv.org/abs/1801.05929} {arXiv:1801.05929
  [cond-mat.mtrl-sci]} \BibitemShut {NoStop}%
\bibitem [{\citenamefont {Liang}\ \emph {et~al.}(2018)\citenamefont {Liang},
  \citenamefont {Lin}, \citenamefont {Kushwaha}, \citenamefont {Xing},
  \citenamefont {Ni}, \citenamefont {Cava},\ and\ \citenamefont {Ong}}]{Ong18}%
  \BibitemOpen
  \bibfield  {author} {\bibinfo {author} {\bibfnamefont {S.}~\bibnamefont
  {Liang}}, \bibinfo {author} {\bibfnamefont {J.}~\bibnamefont {Lin}}, \bibinfo
  {author} {\bibfnamefont {S.}~\bibnamefont {Kushwaha}}, \bibinfo {author}
  {\bibfnamefont {J.}~\bibnamefont {Xing}}, \bibinfo {author} {\bibfnamefont
  {N.}~\bibnamefont {Ni}}, \bibinfo {author} {\bibfnamefont {R.~J.}\
  \bibnamefont {Cava}}, \ and\ \bibinfo {author} {\bibfnamefont {N.~P.}\
  \bibnamefont {Ong}},\ }\href {\doibase 10.1103/PhysRevX.8.031002} {\bibfield
  {journal} {\bibinfo  {journal} {Phys. Rev. X}\ }\textbf {\bibinfo {volume}
  {8}},\ \bibinfo {pages} {031002} (\bibinfo {year} {2018})}\BibitemShut
  {NoStop}%
\bibitem [{\citenamefont {Altland}\ and\ \citenamefont
  {Bagrets}(2016)}]{Altland15}%
  \BibitemOpen
  \bibfield  {author} {\bibinfo {author} {\bibfnamefont {A.}~\bibnamefont
  {Altland}}\ and\ \bibinfo {author} {\bibfnamefont {D.}~\bibnamefont
  {Bagrets}},\ }\href {\doibase 10.1103/PhysRevB.93.075113} {\bibfield
  {journal} {\bibinfo  {journal} {Phys. Rev. B}\ }\textbf {\bibinfo {volume}
  {93}},\ \bibinfo {pages} {075113} (\bibinfo {year} {2016})}\BibitemShut
  {NoStop}%
\bibitem [{\citenamefont {Ma}\ \emph {et~al.}(2017)\citenamefont {Ma},
  \citenamefont {Xu}, \citenamefont {Chan}, \citenamefont {Zhang},
  \citenamefont {Chang}, \citenamefont {Lin}, \citenamefont {Xie},
  \citenamefont {Palacios}, \citenamefont {Lin}, \citenamefont {Jia},
  \citenamefont {Lee}, \citenamefont {Jarillo-Herrero},\ and\ \citenamefont
  {Gedik}}]{Gedik17}%
  \BibitemOpen
  \bibfield  {author} {\bibinfo {author} {\bibfnamefont {Q.}~\bibnamefont
  {Ma}}, \bibinfo {author} {\bibfnamefont {S.-Y.}\ \bibnamefont {Xu}}, \bibinfo
  {author} {\bibfnamefont {C.-K.}\ \bibnamefont {Chan}}, \bibinfo {author}
  {\bibfnamefont {C.-L.}\ \bibnamefont {Zhang}}, \bibinfo {author}
  {\bibfnamefont {G.}~\bibnamefont {Chang}}, \bibinfo {author} {\bibfnamefont
  {Y.}~\bibnamefont {Lin}}, \bibinfo {author} {\bibfnamefont {W.}~\bibnamefont
  {Xie}}, \bibinfo {author} {\bibfnamefont {T.}~\bibnamefont {Palacios}},
  \bibinfo {author} {\bibfnamefont {H.}~\bibnamefont {Lin}}, \bibinfo {author}
  {\bibfnamefont {S.}~\bibnamefont {Jia}}, \bibinfo {author} {\bibfnamefont
  {P.~A.}\ \bibnamefont {Lee}}, \bibinfo {author} {\bibfnamefont
  {P.}~\bibnamefont {Jarillo-Herrero}}, \ and\ \bibinfo {author} {\bibfnamefont
  {N.}~\bibnamefont {Gedik}},\ }\href {http://dx.doi.org/10.1038/nphys4146}
  {\bibfield  {journal} {\bibinfo  {journal} {Nature Physics}\ }\textbf
  {\bibinfo {volume} {13}},\ \bibinfo {pages} {842 EP } (\bibinfo {year}
  {2017})}\BibitemShut {NoStop}%
\bibitem [{\citenamefont {Tabert}\ and\ \citenamefont
  {Carbotte}(2016)}]{Carbotte_optical}%
  \BibitemOpen
  \bibfield  {author} {\bibinfo {author} {\bibfnamefont {C.~J.}\ \bibnamefont
  {Tabert}}\ and\ \bibinfo {author} {\bibfnamefont {J.~P.}\ \bibnamefont
  {Carbotte}},\ }\href {\doibase 10.1103/PhysRevB.93.085442} {\bibfield
  {journal} {\bibinfo  {journal} {Phys. Rev. B}\ }\textbf {\bibinfo {volume}
  {93}},\ \bibinfo {pages} {085442} (\bibinfo {year} {2016})}\BibitemShut
  {NoStop}%
\bibitem [{\citenamefont {Ahn}\ \emph {et~al.}(2017)\citenamefont {Ahn},
  \citenamefont {Mele},\ and\ \citenamefont {Min}}]{Mele_optical}%
  \BibitemOpen
  \bibfield  {author} {\bibinfo {author} {\bibfnamefont {S.}~\bibnamefont
  {Ahn}}, \bibinfo {author} {\bibfnamefont {E.~J.}\ \bibnamefont {Mele}}, \
  and\ \bibinfo {author} {\bibfnamefont {H.}~\bibnamefont {Min}},\ }\href
  {\doibase 10.1103/PhysRevB.95.161112} {\bibfield  {journal} {\bibinfo
  {journal} {Phys. Rev. B}\ }\textbf {\bibinfo {volume} {95}},\ \bibinfo
  {pages} {161112} (\bibinfo {year} {2017})}\BibitemShut {NoStop}%
\bibitem [{\citenamefont {Roy}\ \emph {et~al.}(2016)\citenamefont {Roy},
  \citenamefont {Juri{\v c}i{\'c}},\ and\ \citenamefont
  {Das~Sarma}}]{Roy_optical}%
  \BibitemOpen
  \bibfield  {author} {\bibinfo {author} {\bibfnamefont {B.}~\bibnamefont
  {Roy}}, \bibinfo {author} {\bibfnamefont {V.}~\bibnamefont {Juri{\v
  c}i{\'c}}}, \ and\ \bibinfo {author} {\bibfnamefont {S.}~\bibnamefont
  {Das~Sarma}},\ }\href {http://dx.doi.org/10.1038/srep32446} {\bibfield
  {journal} {\bibinfo  {journal} {Scientific Reports}\ }\textbf {\bibinfo
  {volume} {6}},\ \bibinfo {pages} {32446 EP } (\bibinfo {year}
  {2016})}\BibitemShut {NoStop}%
\bibitem [{\citenamefont {Sun}\ and\ \citenamefont {Wang}(2017)}]{Sun_optical}%
  \BibitemOpen
  \bibfield  {author} {\bibinfo {author} {\bibfnamefont {Y.}~\bibnamefont
  {Sun}}\ and\ \bibinfo {author} {\bibfnamefont {A.-M.}\ \bibnamefont {Wang}},\
  }\href {\doibase 10.1103/PhysRevB.96.085147} {\bibfield  {journal} {\bibinfo
  {journal} {Phys. Rev. B}\ }\textbf {\bibinfo {volume} {96}},\ \bibinfo
  {pages} {085147} (\bibinfo {year} {2017})}\BibitemShut {NoStop}%
\bibitem [{\citenamefont {{Neubauer}}\ \emph {et~al.}(2018)\citenamefont
  {{Neubauer}}, \citenamefont {{Yaresko}}, \citenamefont {{Li}}, \citenamefont
  {{L{\"o}hle}}, \citenamefont {{H{\"u}bner}}, \citenamefont {{Schilling}},
  \citenamefont {{Shekhar}}, \citenamefont {{Felser}}, \citenamefont
  {{Dressel}},\ and\ \citenamefont {{Pronin}}}]{Felser_optical}%
  \BibitemOpen
  \bibfield  {author} {\bibinfo {author} {\bibfnamefont {D.}~\bibnamefont
  {{Neubauer}}}, \bibinfo {author} {\bibfnamefont {A.}~\bibnamefont
  {{Yaresko}}}, \bibinfo {author} {\bibfnamefont {W.}~\bibnamefont {{Li}}},
  \bibinfo {author} {\bibfnamefont {A.}~\bibnamefont {{L{\"o}hle}}}, \bibinfo
  {author} {\bibfnamefont {R.}~\bibnamefont {{H{\"u}bner}}}, \bibinfo {author}
  {\bibfnamefont {M.~B.}\ \bibnamefont {{Schilling}}}, \bibinfo {author}
  {\bibfnamefont {C.}~\bibnamefont {{Shekhar}}}, \bibinfo {author}
  {\bibfnamefont {C.}~\bibnamefont {{Felser}}}, \bibinfo {author}
  {\bibfnamefont {M.}~\bibnamefont {{Dressel}}}, \ and\ \bibinfo {author}
  {\bibfnamefont {A.~V.}\ \bibnamefont {{Pronin}}},\ }\href@noop {} {\bibfield
  {journal} {\bibinfo  {journal} {ArXiv e-prints}\ } (\bibinfo {year}
  {2018})},\ \Eprint {http://arxiv.org/abs/1803.09708} {arXiv:1803.09708
  [cond-mat.mes-hall]} \BibitemShut {NoStop}%
\bibitem [{\citenamefont {Panfilov}\ \emph {et~al.}(2014)\citenamefont
  {Panfilov}, \citenamefont {Burkov},\ and\ \citenamefont
  {Pesin}}]{Panfilov14}%
  \BibitemOpen
  \bibfield  {author} {\bibinfo {author} {\bibfnamefont {I.}~\bibnamefont
  {Panfilov}}, \bibinfo {author} {\bibfnamefont {A.~A.}\ \bibnamefont
  {Burkov}}, \ and\ \bibinfo {author} {\bibfnamefont {D.~A.}\ \bibnamefont
  {Pesin}},\ }\href {\doibase 10.1103/PhysRevB.89.245103} {\bibfield  {journal}
  {\bibinfo  {journal} {Phys. Rev. B}\ }\textbf {\bibinfo {volume} {89}},\
  \bibinfo {pages} {245103} (\bibinfo {year} {2014})}\BibitemShut {NoStop}%
\bibitem [{\citenamefont {Zhou}\ \emph {et~al.}(2015)\citenamefont {Zhou},
  \citenamefont {Chang},\ and\ \citenamefont {Xiao}}]{Xiao15}%
  \BibitemOpen
  \bibfield  {author} {\bibinfo {author} {\bibfnamefont {J.}~\bibnamefont
  {Zhou}}, \bibinfo {author} {\bibfnamefont {H.-R.}\ \bibnamefont {Chang}}, \
  and\ \bibinfo {author} {\bibfnamefont {D.}~\bibnamefont {Xiao}},\ }\href
  {\doibase 10.1103/PhysRevB.91.035114} {\bibfield  {journal} {\bibinfo
  {journal} {Phys. Rev. B}\ }\textbf {\bibinfo {volume} {91}},\ \bibinfo
  {pages} {035114} (\bibinfo {year} {2015})}\BibitemShut {NoStop}%
\end{thebibliography}%
\end{document}